\documentclass[footinbib, aps,pra,reprint, 10pt,twocolumn,showpacs, superscriptaddress]{revtex4-1}
\usepackage{amsfonts,amsmath,amssymb}
\usepackage{blindtext}
\usepackage{graphicx}
\usepackage[pdfstartview=FitH,colorlinks=true,linkcolor=blue,citecolor=blue,urlcolor=blue]{hyperref}
\usepackage{tikz}
\usepackage{verbatim}
\usepackage{comment}

\usepackage{times}

\usepackage[english]{babel}
\usepackage{latexsym}
\usepackage{graphics}
\usepackage{subfigure}
\usepackage{epsfig}
\usepackage{color}
\usepackage{braket}
\usepackage[T1]{fontenc}
\usepackage[latin9]{inputenc}
\usepackage{amstext}
\usepackage{esint}
\usepackage[normalem]{ulem}   

\newcommand{\beq}{\begin{equation}}
\newcommand{\eeq}{\end{equation}}

\begin{document}

\title{Thermal instability, evaporation and thermodynamics of one-dimensional liquids in weakly-interacting Bose-Bose mixtures}

\author{Giulia De Rosi}
\email{giulia.de.rosi@upc.edu}
\affiliation{Departament de F\'isica, Universitat Polit\`ecnica de Catalunya, Campus Nord B4-B5, 08034 Barcelona, Spain}

\author{Grigori E. Astrakharchik}
\email{grigori.astrakharchik@upc.edu}
\affiliation{Departament de F\'isica, Universitat Polit\`ecnica de Catalunya, Campus Nord B4-B5, 08034 Barcelona, Spain}

\author{Pietro Massignan}
\email{pietro.massignan@upc.edu}
\affiliation{Departament de F\'isica, Universitat Polit\`ecnica de Catalunya, Campus Nord B4-B5, 08034 Barcelona, Spain}

\date{\today}

\begin{abstract}
We study the low-temperature thermodynamics of weakly-interacting uniform liquids in one-dimensional attractive Bose-Bose mixtures.~The Bogoliubov approach is used to simultaneously describe quantum and thermal fluctuations. 
First, we investigate in detail two different thermal mechanisms driving the liquid-to-gas transition, the dynamical instability and the evaporation, and we draw the phase diagram. Then, we compute the main thermodynamic quantities of the liquid, such as the chemical potential, the Tan's contact, the adiabatic sound velocity and the specific heat at constant volume. The strong dependence of the thermodynamic quantities on the temperature may be used as a precise temperature probe for experiments on quantum liquids. 
\end{abstract}

\maketitle

\section{Introduction}

Fluids are traditionally classified as gases or liquids depending on the possibility of being self-bound. 
Gases have a positive pressure, so that in absence of a confining container these expand until they occupy the whole available space. 
Liquids are normally much denser than gases and have an equilibrium density fixed by the zero-pressure condition, or, equivalently, by minimizing the free-energy per particle \cite{Landau2013fluid}.  
The equilibrium density is obtained balancing a long-distance attraction (typically of van der Waals type) and a short-range repulsion.
Liquids are therefore self-bound, and the uniform state can easily break down to form small droplets. 
As the temperature is raised, the thermal contribution to the kinetic energy plays against the attractive interactions and makes a liquid less self-bound.
Once a critical temperature is reached, the liquid evaporates into a gas.

Quantum effects play a key role at very low temperatures, where the quantum contribution to the kinetic energy is significant.
While most liquids solidify upon cooling at atmospheric pressure, helium remains liquid even at absolute zero temperature due to the large zero-point motion caused by its small atomic mass.
Understanding the quantum properties of liquid helium is challenging, because it is a dense system with strong interactions, so that perturbative theories cannot be directly applied and a careful modeling of the short-range details of the potential is needed \cite{Boronat1998, Casulleras2000, Boronat2001, Dalfovo2001, Toennies2001, Volovik2003, Diallo2012}. 

A novel and very different class of quantum liquids emerged in recent studies with ultracold atomic gases \cite{Bottcher2021, Luo2021}. 
Ultracold dilute liquids have been experimentally created and investigated in three-dimensional (3D) samples with anisotropic dipolar interactions \cite{Kadau2016, Ferrier-Barbut2016, Wachtler2016, Saito2016, Schmitt2016, FerrierBarbut2016, Chomaz2016, Bottcher2019} and in Bose-Bose mixtures with isotropic contact interactions \cite{Petrov2015, Cabrera2018, Semeghini2018}. 
These liquids exhibit exceptionally low densities (eight orders of magnitude lower than helium's) and temperatures (nine orders of magnitude lower than the freezing point of most classical liquids).
Ultracold droplets can be accurately described by microscopic models which are much simpler than the ones needed for liquid helium. 
In fact, droplets are formed in the weakly-interacting regime, so that perturbative theories can be used.
An ultradilute liquid forms when the interaction strengths are tuned in such a way that the dominant contribution to the energy (the mean-field part) is close to vanishing. 
Under these conditions, the importance of beyond-mean-field quantum fluctuations (which are usually subleading) is greatly enhanced, and these can provide the balancing mechanism needed to obtain a liquid. 
Quantum droplets are therefore stable due to genuine many-body effects.
Such liquids have a number of advantages as there is an exquisite experimental control over interactions, temperature, spatial dimension, trapping, and relative populations in mixtures. 
As such, they can model other complex and very different many-body systems, including liquid helium. 
The evaporation leads to the reduction of the number of particles in the liquid as discussed for the zero temperature case in Refs.~\cite{Petrov2015, Cikojevic2017, Ancilotto2018}, although it cannot be clearly distinguished from three-body losses in current experiments \cite{Cabrera2018, Semeghini2018}. 

These quantum liquids have been predicted to exist also in one-dimensional (1D) Bose-Bose mixtures \cite{Petrov2016, Parisi2019, Hu2020}, where beyond-mean-field effects are greatly enhanced, and three-body collisions strongly reduced \cite{Tolra2004,Lavoine2021}. A 1D configuration should therefore allow for a more precise characterization of the evaporation process. 
First experiments with quantum droplets in confined geometries have been recently reported \cite{Cheiney2018}. 
Differently from the three-dimensional case, the stability of 1D liquids requires a net mean-field repulsion which competes with an effective attraction provided by quantum fluctuations \cite{Petrov2016}.
One-dimensional liquids generated a significant interest, and recent works explored their zero temperature phase diagram \cite{Parisi2019} and dynamics \cite{Astrakharchik2018, Parisi2020, Tylutki2020}.

However, the experiments are inherently performed at finite temperature \cite{Cabrera2018, Cheiney2018}, and an important open problem is the understanding of thermal effects in quantum liquids.
These systems in particular require novel temperature probes \cite{Bottcher2021}, because the standard technique for thermometry in ultracold gases (time-of-flight expansion) cannot obviously be applied to self-bound systems. 
One should therefore resort to {\it in-situ} techniques, such as the ones exploited in recent ultracold-gas experiments to measure with high-precision several thermodynamic quantities. 
These include the sound velocity, the free energy, the specific heat at constant volume, the Tan's contact parameter and the chemical potential \cite{Andrews1997, Joseph2007, Ho2009, Nascimbene2010b, Ku2012, Hoinka2013, Salces-Carcoba2018}.

The enhanced stability of one-dimensional mixtures opens the intriguing perspective of investigating the influence of temperature on liquids, whose understanding is very limited at the moment. 
A thermal instability of quantum liquids, inducing the transition to the gas phase, has been analyzed from the perspective of the pairing theory \cite{Wang2020} and of the non-interacting phononic excitations description \cite{Ota2020}. 
As the temperature is increased, however, higher quasi-particle momentum states get populated and the deviation of the excitation spectrum from the simple linear phononic behavior are expected to become important \cite{Meinert2015, Fabbri2015, DeRosi2019}. 
To our knowledge, no previous studies of the evaporation in 1D liquids, both at zero and finite temperature, were reported.

In this work, we study weakly-interacting uniform liquids formed in one-dimensional Bose-Bose mixtures at finite temperature.
We provide a comprehensive description of their thermodynamic quantities at finite temperature, and we study in detail the two mechanisms which rule the thermal liquid-to-gas transition: the dynamical instability and the evaporation, which occur respectively in the bulk and at the surface. 
To perform our study we employ the Bogoliubov (BG) theory, which describes the low-temperature thermodynamic behavior of the system in terms of non-interacting BG quasi-particles. 
This method has been previously successfully applied to a single-component 1D Bose gas at finite temperature, showing an excellent agreement with exact Bethe ansatz techniques of all the thermodynamic quantities \cite{DeRosi2019}. In this paper, we generalize that approach to bosonic mixtures. 
The BG theory takes into account the nonlinearity of the excitation spectrum, thus going beyond simpler non-interacting phonon models \cite{DeRosi2017, DeRosi2019, Ota2020}. 
By doing so it is possible to include the \textit{beyond-mean-field} quantum fluctuations even in the thermal effects. 
The strong dependence of the dynamical instability and the evaporation on the coupling constants, combined with the fine tunability of the interactions ensured by Fano-Feshbach resonances \cite{DErrico2007, Chin2010, Tanzi2018}, can be employed to measure the temperature in quantum liquids with unprecedented precision. 
Finally, our theoretical predictions constitute a fundamental benchmark for future in-situ measurements of key thermodynamic quantities. 

The paper is organized as follows. 
In Sec.~\ref{Sec:Model} we present the system under investigation, 
introduce the thermodynamic quantities of interest, and derive a series of exact relations linking them. 
In Sec.~\ref{Sec:liquids zero T} we calculate the ground-state properties of the liquid at zero temperature. 
In Sec.~\ref{Sec:Low T transition} we discuss the phase diagram of the system at finite temperature, investigate the thermal liquid-gas transition and compute the critical temperatures of both the dynamical instability and the evaporation. 
In Sec.~\ref{Sec:low-T liquid} we calculate several thermodynamic quantities of the liquid at finite temperature. 
In Sec.~\ref{Sec:experiments} we discuss the experimental relevance of our predictions. 
In Sec.~\ref{Sec:conclusions} we draw our conclusions and we present  future perspectives of our results. 
In App.~\ref{Sec:thermodynamic relations}-\ref{Sec:Inverse adiabatic compressibility} we provide details about the derivation of the exact thermodynamic relations and the adiabatic sound velocity, respectively. 

\section{Model}
\label{Sec:Model}
Throughout this paper we consider a one-dimensional uniform mixture of two bosonic components with pairwise contact interactions, whose Hamiltonian reads:
\begin{multline}
\label{Eq:H}
H = \sum_{\sigma = 1}^2 \left[ - \frac{\hbar^2}{2m}\sum_{i = 1}^{N_\sigma} \frac{\partial^2}{\partial x_i^2} + g\sum_{i > j}^{N_\sigma}\delta(x_i - x_j) \right] + \\ g_{12} \sum_{i > j}^{N_1, N_2} \delta(x_i - x_j).
\end{multline}
Here $N_\sigma$ is the number of atoms in component $\sigma = 1,2$, and to simplify the treatment we restrict ourselves to study balanced mixtures with $N_1 = N_2 = N/2$. 
The total linear density is therefore $n = n_1 + n_2 = N/L$, where $L$ is the length of the system. 
All atoms have the same mass $m$, and experience the same intra-species repulsive interactions controlled by the coupling constant $g = - 2 \hbar^2/(m a) > 0$, where  $a<0$ is the 1D intra-species $s$-wave scattering length. 
We consider attractive inter-species interaction with coupling constant $g_{12} = - 2 \hbar^2/(m a_{12}) < 0$ depending on the 1D inter-species $s$-wave scattering length $a_{12} > 0$.
In the following, we will focus on the regime close to the mean-field instability point $|g_{12}| \lesssim g$, where a weakly-interacting quantum liquid emerges at zero temperature   \cite{Petrov2016, Parisi2019}. 
In this regime the perturbative Bogoliubov theory can be safely applied. 
While in three dimensions this approach describes the low-density limit, a peculiar feature of one-dimensional geometry is that the weakly-interacting regime corresponds to the high-density limit $n |a| \gg 1$ and $n a_{12} \gg 1$. 

Within the canonical ensemble, the complete thermodynamics of the system can be obtained from the Helmholtz free energy density $\mathcal{A} = \mathcal{E}-T\mathcal{S}$, where $\mathcal{E}$ is the internal energy density and $\mathcal{S}$ is the entropy density. 
The energy cost of adding a single particle to the system, or chemical potential, is given by
\begin{equation}
\label{Eq:mu}
\mu = \left( \frac{\partial \mathcal{A}}{\partial n}    \right)_{T, a, a_{12}} .
\end{equation}  

Another key thermodynamic quantity is the pressure,
\begin{equation}
\label{Eq:P}
P = n \mu - \mathcal{A} .
\end{equation}
The pressure is always positive in a gas state. 
Liquids can even sustain a negative pressure, when its value is not too large.

A special property of a many-body self-bound state at zero pressure and temperature is that the total energy is related to the chemical potential as $E=\mu N$, which might be confronted with the relation $E=\mu N/2$ holding for a single-component mean-field weakly repulsive Bose gas \cite{Lieb1963}. Thus, the condition of formation of a liquid at $T=0$ and $P=0$ is negative energy or chemical potential.
The chemical potential at zero temperature sets the particle emission threshold $- \mu$ and determines the rate of evaporation of atoms from a liquid, which at zero temperature is ruled by quantum fluctuations only \cite{Petrov2015}.
At finite temperature, thermal fluctuations start to play an important role when the thermal energy becomes larger than the particle emission threshold.

A central role is also played by the entropy density 
\begin{equation}
\label{Eq:S}
\mathcal{S} = - \left(\frac{\partial \mathcal{A}}{\partial T}    \right)_{a, a_{12}, n}.
\end{equation}
Indeed, expressing the pressure $P$ in terms of the density $n$ and the entropy per particle $\bar{s} = \mathcal{S}/n$ yields the equation of state $P(n,\bar{s})$.
 
The low-lying collective motion of the system propagate at the adiabatic sound velocity $v$, which is obtained from the equation of state through the relation  \cite{DeRosi2015, DeRosi2016}
\begin{equation}
\label{Eq:v}
m v^2 = \left( \frac{\partial P}{\partial n}  \right)_{\bar{s}}.
\end{equation}

The system is dynamically stable provided its inverse isothermal compressibility 
 \begin{equation}
 \label{Eq:kT}
 \kappa_T^{-1} = \left(  \frac{\partial^2 \mathcal{A}}{\partial n^2} \right)_{T, a, a_{12}} 
 \end{equation}
is positive.

The specific heat density at constant volume (or length $L$, in one dimension) is
\begin{equation}
\label{Eq:specific heat}
C_L = \left(  \frac{\partial \mathcal{E}}{\partial T}  \right)_{a, a_{12}, n} .
\end{equation}

Another important thermodynamic quantity characterizing ultracold atoms with zero-range interactions is the Tan's contact parameter. It provides a number of useful exact relations linking the equation of state, the pressure, the total and the interaction energies, and the short-distance (large momentum and high-frequency) properties of the correlation functions  \cite{Barth2011, Wild2012, Olshanii2003, Tan2008, Tan22008, Tan32008}. The intra-species Tan's contact parameter density is  \cite{Braaten2011, Yao2018}
\begin{equation}
\label{Eq:contact}
\mathcal{C_+} =  \frac{4 m }{\hbar^2} \left(  \frac{\partial \mathcal{A}}{\partial a}   \right)_{T, a_{12}, n}.
\end{equation}
The inter-species Tan's contact parameter density $\mathcal{C}_{-}$ is similarly defined, interchanging the role of $a$ and $a_{12}$. 

 Simple considerations based on scale invariance  \cite{FetterBook, Barth2011, Patu2017} lead to a series of exact thermodynamic relations linking the aforementioned quantities, which hold for arbitrary temperature and strength of (contact) interactions:
\begin{equation}
\label{Eq:thermodynamic identities}
- \frac{\hbar^2}{4 m} \left(\mathcal{C} a + \mathcal{C}_{12} a_{12}   \right) = 3 \mathcal{A} + 2 T \mathcal{S} - n \mu = 2 \mathcal{E} - P.
\end{equation}
Their complete derivation is provided in Appendix~\ref{Sec:thermodynamic relations}.

A relation connecting the Tan's contact, the pressure and the energy, which is similar to Eq.~\eqref{Eq:thermodynamic identities}, holds in two dimensional ultracold gases. 
In that case, it is a consequence of the quantum anomaly effect, i.e.~a quantum-mechanical symmetry breaking, since the contact parameter modifies the scale-invariant energy-pressure relation. 
In Bose gases, first theoretical investigations revealed a universality of the breathing mode frequency in trapped gas showing a scaling symmetry  \cite{Pitaevskii1997}. 
Such symmetry actually breaks under quantization resulting in a small shift away from the scale-invariant value of the breathing mode frequency  \cite{Olshanii2010}.
Recently, the quantum anomaly has been observed in experiments with two-dimensional Fermi gases \cite{Holten2018, Peppler2018}.

\section{Liquids at zero temperature}

\label{Sec:liquids zero T}
At the mean-field level, a weakly-interacting Bose-Bose mixture at zero temperature has a ground-state energy density given by
\begin{equation}
\label{Eq:E0MF}
\mathcal{E}_{0,{\rm mf}} = g n_1^2/2 + g n_2^2/2 - |g_{12}|n_1 n_2=nmc_-^2/2.
\end{equation}
Diagonalizing this quadratic form yields two long-wavelength phononic excitations with sound velocities $c_-<c_+$ given by
\begin{equation}
\label{Eq:cpm2}
c_\pm^2 = \frac{n}{m} \frac{g \pm |g_{12}| }{2}.
\end{equation}
Within the mean-field description and for equal populations ($n_1=n_2 = n/2$), the system is stable provided that $|g_{12}|<g$, while it undergoes a collapse to a soliton for stronger interspecies attraction. 

At the beyond mean-field level, a standard analysis  \cite{Pethick2008} shows that the excitation spectrum of the mixture contains two Bogoliubov branches with dispersion relations
\begin{equation}
\label{Eq:BG spectrum}
E_\pm(p) = \sqrt{c_\pm^2 p^2 + \left(  \frac{p^2}{2m} \right)^2} .
\end{equation} 
With attractive inter-species interactions, the soft mode $E_-$ describes ``density'' oscillations where the two mixture components oscillate in phase, while the stiffer mode $E_+>E_-$ corresponds to ``spin'' oscillations where the two components oscillate out of phase. 
The first beyond mean-field correction to the energy, describing quantum fluctuations, is purely attractive in one dimension, and the total ground-state energy density is  \cite{Petrov2016, Parisi2019}:
\begin{equation}
\label{Eq:E0}
\mathcal{E}_0 = \frac{1}{2} n m c^2_- - \frac{2}{3} \frac{m^2}{\pi \hbar}\sum_\pm c^3_\pm,
\end{equation}
where $\sum_\pm x_\pm$ is a shorthand notation for $x_++x_-$. The leading mean-field term $\propto (g-|g_{12}|)n^2$ can be strongly reduced by tuning appropriately the two coupling constants. In this way one can boost the importance of the beyond-mean field corrections, which are usually subleading since they scale as $n^{3/2}$.
The different power-law dependence of the two contributions, together with their opposite signs (provided $|g_{12}|<g$), immediately shows that there exists a specific density at which the energy per particle has a minimum, which corresponds to the equilibrium density of the quantum liquid  \cite{Petrov2016, Parisi2019}.
  
At zero temperature, the free energy coincides with the internal energy, $\mathcal{A} = \mathcal{E}_0$.
The ground-state properties of the liquid state at zero temperature have been studied in Ref.~  \cite{Parisi2019}.
The chemical potential, Eq.~\eqref{Eq:mu}, is:
\begin{equation}
\label{Eq:mu0}
\mu_0 = mc_-^2 - \frac{m^2}{\pi \hbar n} \sum_\pm c_\pm^3 
\end{equation}
and the pressure is $P_0 = n \mu_0 - \mathcal{E}_0$. Using Eq.~\eqref{Eq:v} one may now compute the ``beyond mean-field'' sound velocity,
\begin{equation}
\label{Eq:v0}
v_0 = \sqrt{c_-^2 - \frac{m}{2 \pi \hbar n} \sum_\pm c_\pm^3}.
\end{equation}
which explicitly contains a contribution from quantum fluctuations  \cite{Lieb21963}.

The \textit{equilibrium} density is found by minimizing the ground-state energy per particle $d \left( \mathcal{E}_0/n  \right)/dn = 0$ or, equivalently, by requiring the pressure to vanish, $P_0 = 0$. In this way one finds  \cite{Petrov2016}: 
\begin{equation}
\label{Eq:neq}
n_{\rm eq}\left( T = 0\right) 
= \frac{2}{9} \frac{m}{\pi^2 \hbar^2} \left[ \frac{\sum_\pm  \left(g \pm |g_{12}| \right)^{3/2}}{g - |g_{12}|}\right]^2 .
\end{equation}
At the equilibrium density, the repulsive mean-field and the attractive beyond mean-field contributions balance each other and a stable quantum liquid can be formed.

The inverse isothermal compressibility becomes
\begin{equation} 
\label{Eq:k0}
\kappa_{T}^{-1}\left( T = 0  \right) = \frac{m c_-^2}{n} - \frac{m^2}{2 \pi \hbar n^2} \sum_\pm c_\pm^3.
\end{equation}
Its mean-field term depends only on the density sound velocity $c_-$.~On the other hand, the mean-field term of the inverse magnetic susceptibility \footnote{
The inverse magnetic susceptibility at zero temperature is:
\begin{equation*}
\label{Eq:kM0}
\kappa_{M}^{-1}\left( T = 0  \right) = 
\left(  \frac{\partial^2 \mathcal{E}_0}{\partial \tilde{n}^2} \right)_{\tilde{n} = 0} = \frac{m c_+^2}{n} - 
\frac{2 m^2}{\pi \hbar n^2} \frac{c_-^2 c_+^2}{c_+ + c_-}
\end{equation*}
where $\tilde{n} = n_1 - n_2$.
} is proportional to the spin sound velocity $c_+$  \cite{Stringari2009, Pitaevskii2016}.
In the vicinity of the mean-field collapse $g \sim |g_{12}|$ (i.e.~$c_- \sim 0$), the inverse isothermal compressibility is negative, signaling the instability of the system. In three dimensions, the isothermal compressibility is instead positive for $c_- \sim 0$, so that a stable fluid governed only by quantum fluctuations can exist  \cite{Jorgensen2018}.
The isothermal compressibility $\kappa_T$ diverges at the {\it spinodal} density $n_{\rm sp} = 9 n_{\rm eq}/16$  \cite{Petrov2016}. For $n > n_{\rm sp}$, the compressibility is positive and the liquid is dynamically stable against local density fluctuations  \cite{Landau2013}. It has a positive or negative pressure depending on whether $n$ is larger or smaller than $n_{\rm eq}$.
At densities below the spinodal point $n < n_{\rm sp}$, the liquid is dynamically unstable and breaks into droplets \cite{Parisi2019}, whose density is approximately equal to the equilibrium value of the uniform phase.

Finally, the zero-temperature intra-species ($\mathcal{C}_+$) and inter-species ($\mathcal{C}_-$) Tan's contact densities are given by:
\begin{equation}
\label{Eq:C0}
\mathcal{C}_\pm \left( T = 0 \right)= \frac{m^4}{2 \hbar^4} \left( c_+^2 \pm c_-^2  \right)^2 \left[ 1 - \frac{2 m}{\pi \hbar n} \left(c_- \pm c_+  \right)      \right] .
\end{equation}

Throughout this work, we limit ourselves to values of the ratio of coupling constants $|g_{12}|/g \geq 0.7$, where the BG theory proved in excellent agreement with recent zero-temperature Quantum Monte-Carlo calculations  \cite{Parisi2019,Parisi2020}.

\section{Quantum Thermal liquid-gas transition}
\label{Sec:Low T transition}

At temperatures $T/T_d \ll \left(  n |a|\right)^{-1/2} \ll 1$, where $k_B T_d = \hbar^2 n^2/(2m)$ is the quantum degeneracy energy \cite{Kheruntsyan2003}, the thermodynamics of the weakly-interacting Bose-Bose mixture can be understood via BG theory in terms of a gas of non-interacting bosonic quasi-particles. 
The thermal free energy density $\Delta \mathcal{A} = \mathcal{A} - \mathcal{E}_0$ is
\begin{equation}
\label{Eq:thermal A quasi-particles}
\Delta \mathcal{A} = 
k_B T  \sum_{\pm} \int_{-\infty}^{+\infty}
\frac{dp}{2 \pi \hbar} 
\ln\left[ \frac{1}{f\left(E_\pm \right) + 1} \right] ,
\end{equation}
where $f\left(E_\pm\right) = \left(e^{\beta E_\pm}  - 1\right)^{-1}$ is the Bose function and $\beta = (k_B T)^{-1}$.
The thermal free energy density $\Delta\mathcal{A}$ depends on the inter- and intra-species interactions through the sound velocities $c_\pm$ which appear in the BG dispersion relations $E_\pm$, see Eq.~\eqref{Eq:BG spectrum}. 

At very low temperatures $k_B T \ll m c_{-}^2$, the Bose-Bose mixture exhibits features of superfluids  \cite{Astrakharchik2004} with linear phononic excitations (i.e., a Luttinger liquid theory applies). 
In this regime, one can generalize the description of a single component  \cite{DeRosi2017} to mixtures  \cite{Ota2020}, by retaining only the phononic part of the BG dispersions, $E_{\pm}(p) = c_{\pm} |p|$ in Eq.~\eqref{Eq:thermal A quasi-particles}, and obtains the first $O\left(T^2\right)$ thermal correction of $\mathcal{A}$:
\begin{equation}
\label{Eq:A phonons}
\Delta \mathcal{A}_{\rm ph} = - \frac{\pi}{6} \frac{\left( k_B T \right)^2}{\hbar} \sum_\pm \frac{1}{c_\pm} .
\end{equation}
A higher accuracy can be obtained including the non-linear part of the dispersion relation. 
Expansion of the BG spectrum according to
$E_{\pm}(p \ll m c_-) \approx  c_{\pm} |p| \left[ 1 + p^2/\left(8 m^2 c_{\pm}^2\right) \right]$ allows one to compute thermal corrections of order $O(T^4)$ \cite{DeRosi2019}. 
At temperatures $k_B T \gg k_B T_d$, a reliable description of the thermodynamics is instead provided by the Hartree-Fock theory  \cite{DeRosi2019}, which is perturbative in the coupling constants  \cite{Pitaevskii2016}.

The thermal excitations are most effective in exciting the lowest energy mode out of the two branches. 
In Bose-Bose mixtures with repulsive $g_{12} > 0$ interactions in bulk \cite{Ota2019, Hryhorchak2021} and optical lattices \cite{Suthar2017}, the lowest energy mode is the spin one and thermal excitations might induce a magnetic phase separation of the system occurring at the divergence of the magnetic susceptibility  \cite{Ota2019}. 
Instead, in our case, the inter-species interactions are attractive, $g_{12} < 0$, and the density modes are the first thermally excited thereby preventing magnetic phase separation. 
Therefore, from Eq.~\eqref{Eq:thermal A quasi-particles}, we calculate only the thermal contribution of the inverse isothermal compressibility $\Delta \kappa^{-1}_{T} = \kappa_{T}^{-1} -  \kappa_{T}^{-1}\left(T = 0\right)$:
\begin{multline}
\label{Eq:kT general spectrum}
\Delta \kappa_T^{-1} =  -  \frac{1}{4 n^2} \sum_{\pm} \int_{-\infty}^{+\infty} \frac{dp}{2 \pi \hbar} \frac{p^4 c_\pm^4}{E^3_\pm(p)} \times \\
\left[ f\left(E_\pm\right) - \beta \frac{\partial}{\partial \beta}  f\left(E_\pm\right)    \right] .
\end{multline}

\begin{figure}[t]
\begin{center}
\includegraphics[width=\columnwidth,angle=0,clip=]{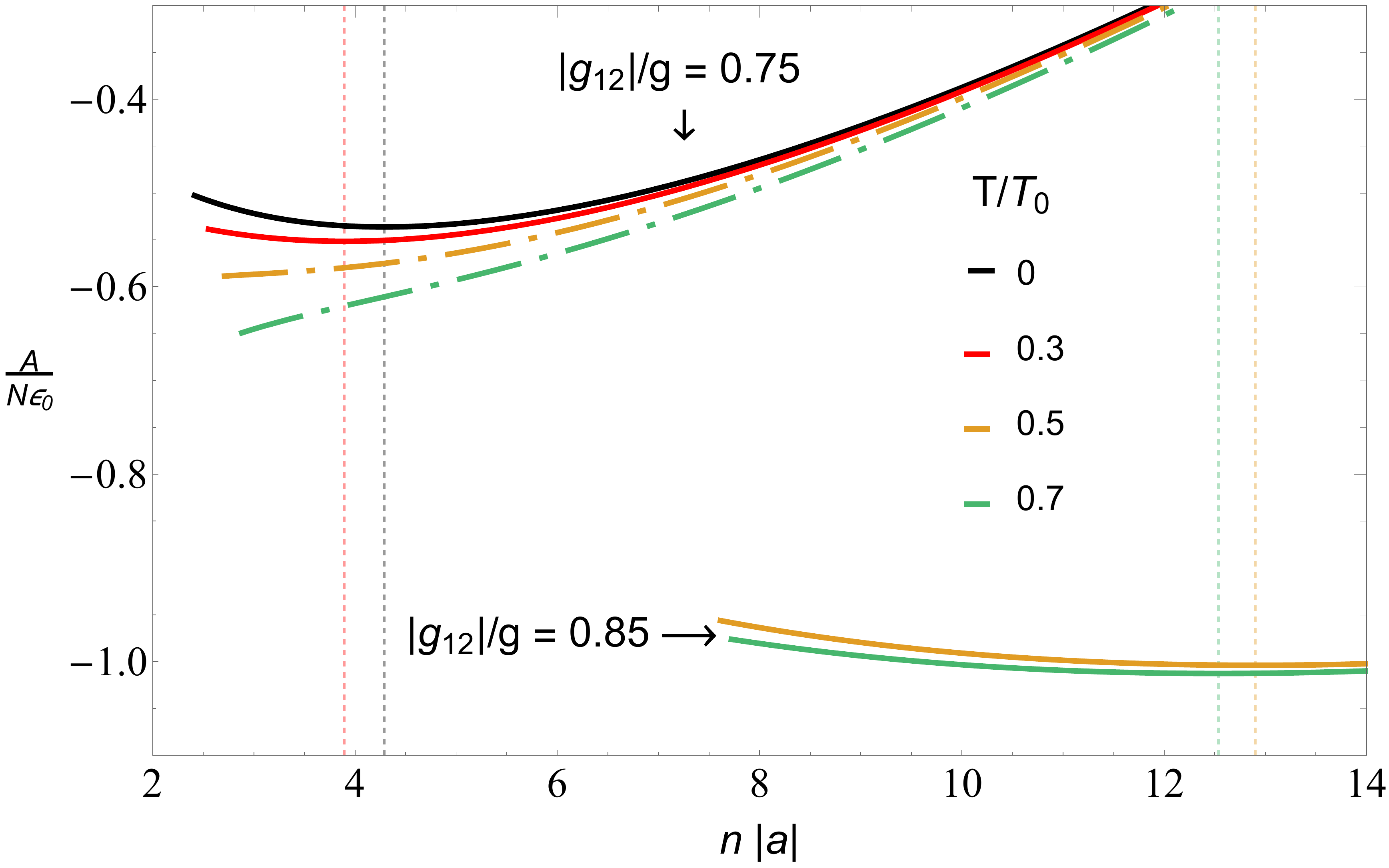}
\caption{
\label{fig:A}
Free energy per particle as a function of the density at different values of temperature in units of $T_0 = \epsilon_0/k_B$, where $\epsilon_0 = \hbar^2/(m |a|^2)$.~The curves are reported in an increasing order of the temperature from low (top) to high (bottom) values.~Upper group of lines are computed for $|g_{12}|/g = 0.75$, while the lower one for $|g_{12}|/g = 0.85$.~Solid lines correspond to the liquid state characterized by a minimum at the equilibrium density $n_{\rm eq}$ (vertical dotted).~Dot-dashed lines have no local minimum, so they yield a gas phase.~Curves are reported for values of densities in the dynamically stable regime. 
}
\end{center}
\end{figure}

Typical results for the free energy per particle $A/N$ are shown in Fig.~\ref{fig:A}. 
All the curves are reported only in the respective dynamically stable regime $n > n_{\rm sp}(T)$, where the spinodal density $n_{\rm sp}$ is the one yielding a vanishing inverse compressibility, $\kappa_T^{-1} = 0$.
The equilibrium densities $n_{\rm eq}(T)$ are given by the extremum condition $d\left(\mathcal{A}/n\right)/d n  = 0$, or from the equivalent relation $P = 0$. 
For sufficiently low temperatures, the free energy per particle admits a local minimum at the equilibrium density $n_{\rm eq}$ (shown with vertical dotted lines), for which the system is a liquid. 
We observe that an increase in the temperature for a fixed value of $|g_{12}|/g$ makes the free-energy per particle more negative and one might think that thermal fluctuations enhance the stability of the liquid. 
Instead, the equilibrium density decreases  \cite{Wang2020} and approaches the spinodal point $n_{\rm eq} \to n_{\rm sp}$, thereby making the liquid unstable. 
Above a critical temperature $T_c$, the free energy per particle becomes a monotonically increasing function of the density (dot-dashed lines), so that the system is in a gas phase. 

\begin{figure}[t]
\begin{center}
\includegraphics[width=\columnwidth,angle=0,clip=]{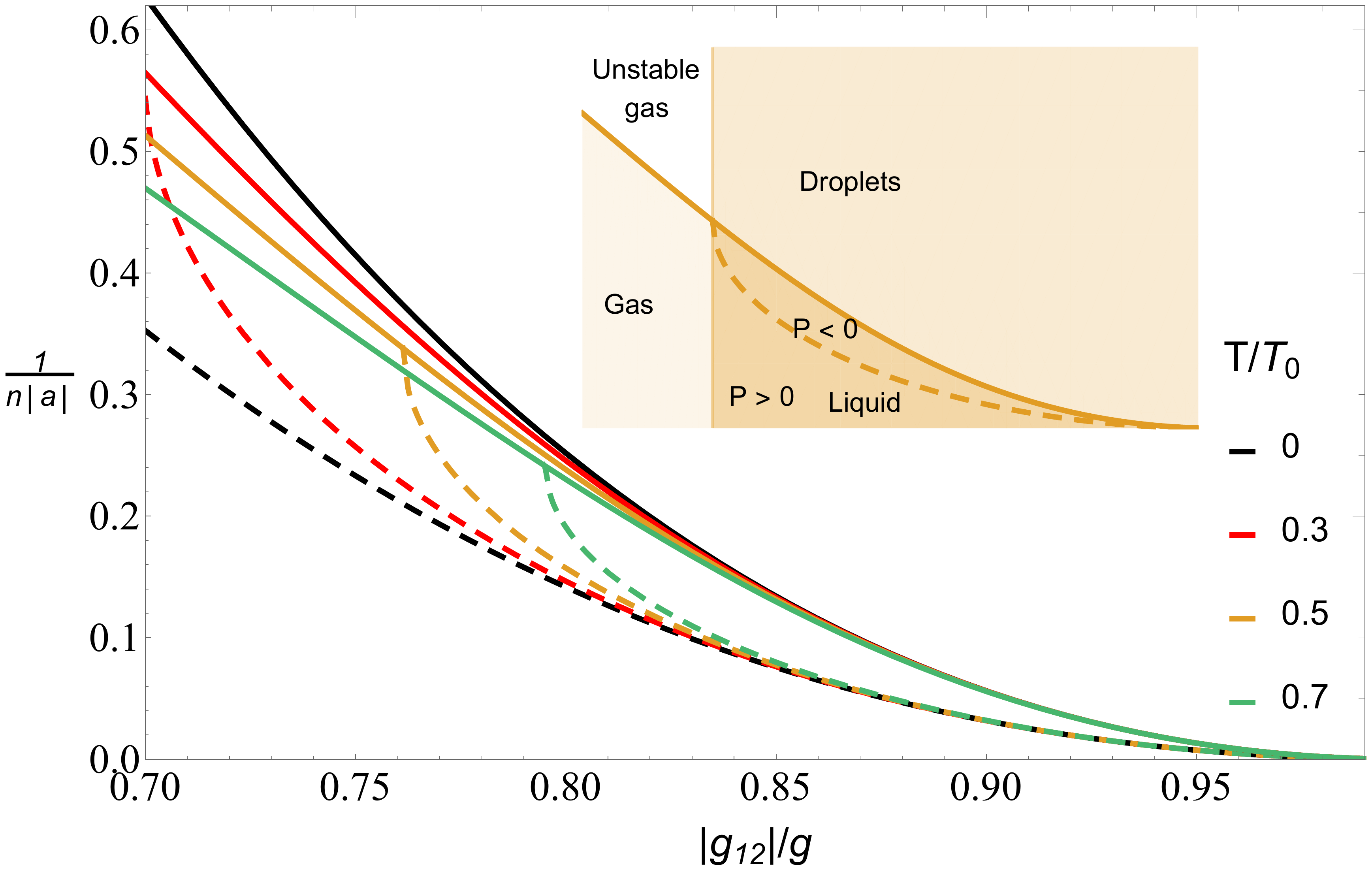}
\caption{
\label{fig:phase diagram}
Phase diagram of 1D weakly-interacting Bose-Bose mixtures at finite temperature.~Spinodal densities (solid) of the mixture and equilibrium densities (dashed) of the liquid are reported.~The various solid (dashed) lines are obtained for increasing (decreasing) values of the temperature, from top to bottom.~The inset shows the typical phase diagram at a fixed temperature. Below the spinodal line, the system is a liquid or a gas depending on whether its free energy per particle has a local minimum or not. In the liquid phase, the pressure is positive or negative depending on whether the density is larger or smaller than the equilibrium value.~Above the spinodal line, both phases are dynamically unstable. 
}
\end{center}
\end{figure}

In Figure~\ref{fig:phase diagram} we show the equilibrium and spinodal densities as dashed and solid lines, respectively, for various values of $|g_{12}|/g$ and inverse density $1/(n|a|)$ at different temperatures. 
This gives the complete phase diagram of the mixture. 
The points at which the dashed and solid lines meet denote the critical interaction strength $(|g_{12}|/g)_{c}$ below which a liquid ceases to exist and transforms into a gas phase. 
For $|g_{12}|/g > (|g_{12}|/g)_{c}$  the liquid survives with its equilibrium density $n_{\rm eq}$. 
By further increasing the temperature, the transition occurs for larger values of $(|g_{12}|/g)_c$, and the gas tends to occupy an increasingly larger portion of the phase diagram. 
Thermal effects are, hence, dominant at smaller values of $|g_{12}|/g$, where they are driven by a larger value of the density sound velocity $c_-$, Eq.~\eqref{Eq:cpm2}, corresponding to the soft mode  \cite{Ota2020}. 

\begin{figure}[t]
\begin{center}
\includegraphics[width=\columnwidth,angle=0,clip=]{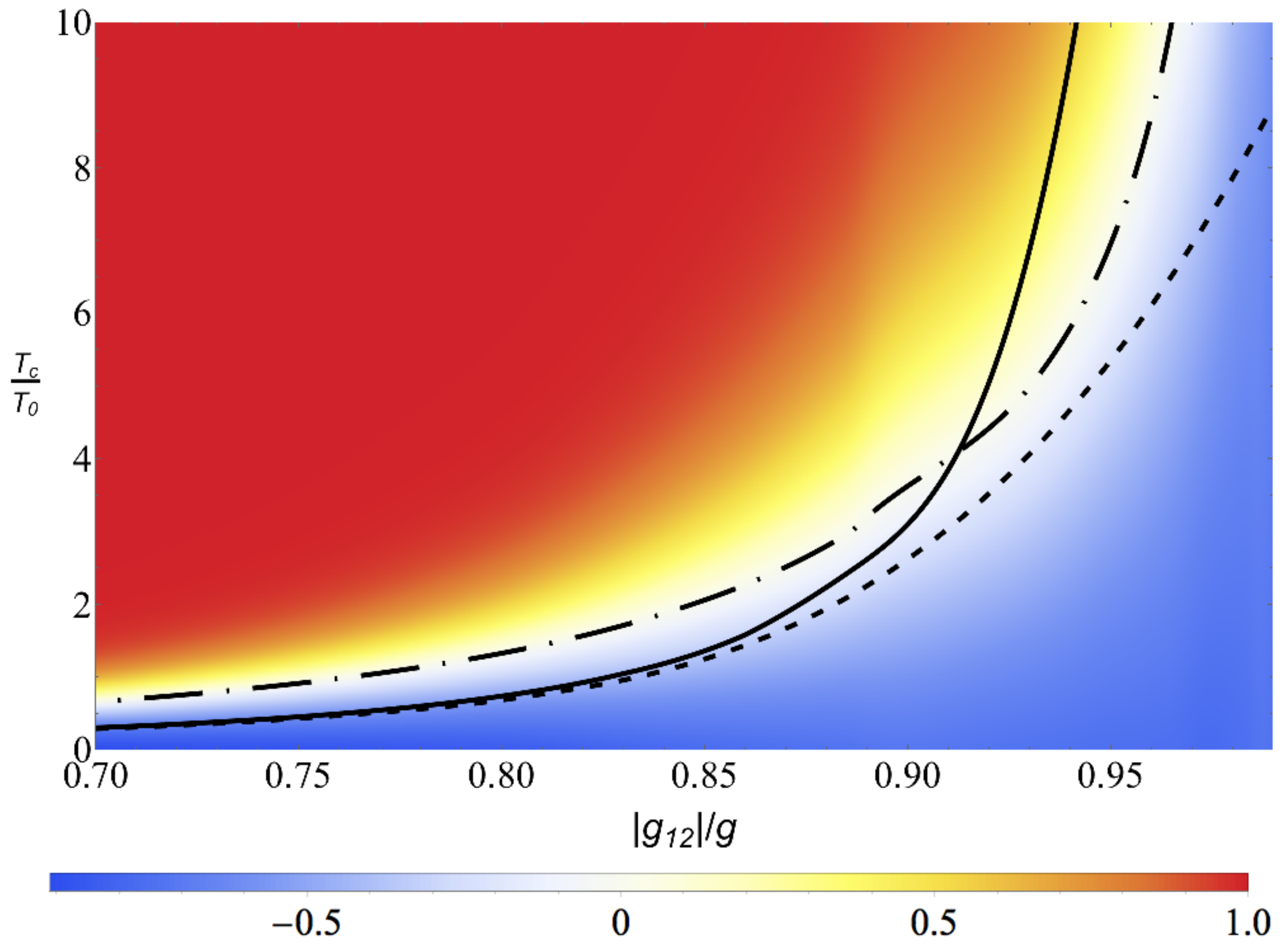}
\caption{Critical temperature $T_c$ of the thermal liquid-gas phase transition as a function of $|g_{12}|/g$.
The threshold temperature of the dynamical instability has been calculated with the full BG dispersion (solid), Eq.~\eqref{Eq:thermal A quasi-particles}, and the phononic spectrum (dashed), Eq.~\eqref{Eq:A phonons}. 
The dot-dashed line corresponds to the typical evaporation temperature $T_{\rm ev}$.
The plot is color-coded using the function $\tanh\left(\frac{T_c -T{\rm ev}}{2\Delta T}\right)$, where $\Delta T = \left(T_{\rm ev} - T_{\rm ev}^\prime \right)$ and $T_{\rm ev}^\prime$ is the solution of the equation $k_B T = - \mu_0\left[n_{\rm eq}(T)\right]$.
In this way, blue and red shadings denote respectively regions of slow and fast evaporation.
}
 \label{fig:Tc}
 \end{center}
\end{figure}
The thermal liquid-gas transition caused by the dynamical instability takes place when the equilibrium density coincides with the spinodal one, so that the minimum in the free energy per particle, characteristic of a liquid state, disappears. The corresponding critical temperature $T_c$ is shown with a solid line in Fig.~\ref{fig:Tc}. 
If instead of the Bogoliubov spectra, one considers only linear density and spin phonons, the critical temperature is significantly underestimated for the large values of $|g_{12}|/g$, where the system is dense. Such prediction, corresponding to Eq.~\eqref{Eq:A phonons}, was first obtained in Ref.~\cite{Ota2020} and is reported with a dashed line in Fig.~\ref{fig:Tc}. In addition to the dynamical instability, thermal fluctuations lead to the evaporation of the liquid. 
The characteristic temperature associated with the evaporation $T_{\rm ev}$ can be estimated from the condition $k_B T/2 = - \mu_0\left[n_{\rm eq}(T)\right]$ (dot-dashed line in Fig.~\ref{fig:Tc}). 
We notice that the solid and dot-dashed lines cross each other at  temperature $T_{\rm cross} \approx 4 T_0$. 
For $T \lesssim T_{\rm cross}$, the solid line is below the dot-dashed one, and the dominant mechanism for the transition is provided by the dynamic instability. For $T \gtrsim T_{\rm cross}$, corresponding to the regime close to the mean-field collapse $|g_{12}|/g \sim 1$, the transition to the gas phase is driven mostly by evaporation. 

At fixed ratio of the interaction strengths, our phase diagram shows that it is possible to create a liquid from a gas by decreasing the temperature. Moreover, the appearance of the dynamical instability or the evaporation is fully controlled by finely tuning the interaction ratio $|g_{12}|/g$ around the crossing point of their critical temperatures. 
Finally, since the results of Fig.~\ref{fig:Tc} have been derived at the equilibrium densities, they apply directly to large (saturated) self-bound droplets, where the central density has reached the equilibrium value of the uniform phase \cite{Petrov2015, Cikojevic2017}.

\section{Low-$T$ thermodynamics of the liquid}
\label{Sec:low-T liquid}

\begin{figure}[t]
\begin{center}
\includegraphics[width=\columnwidth,angle=0,clip=]{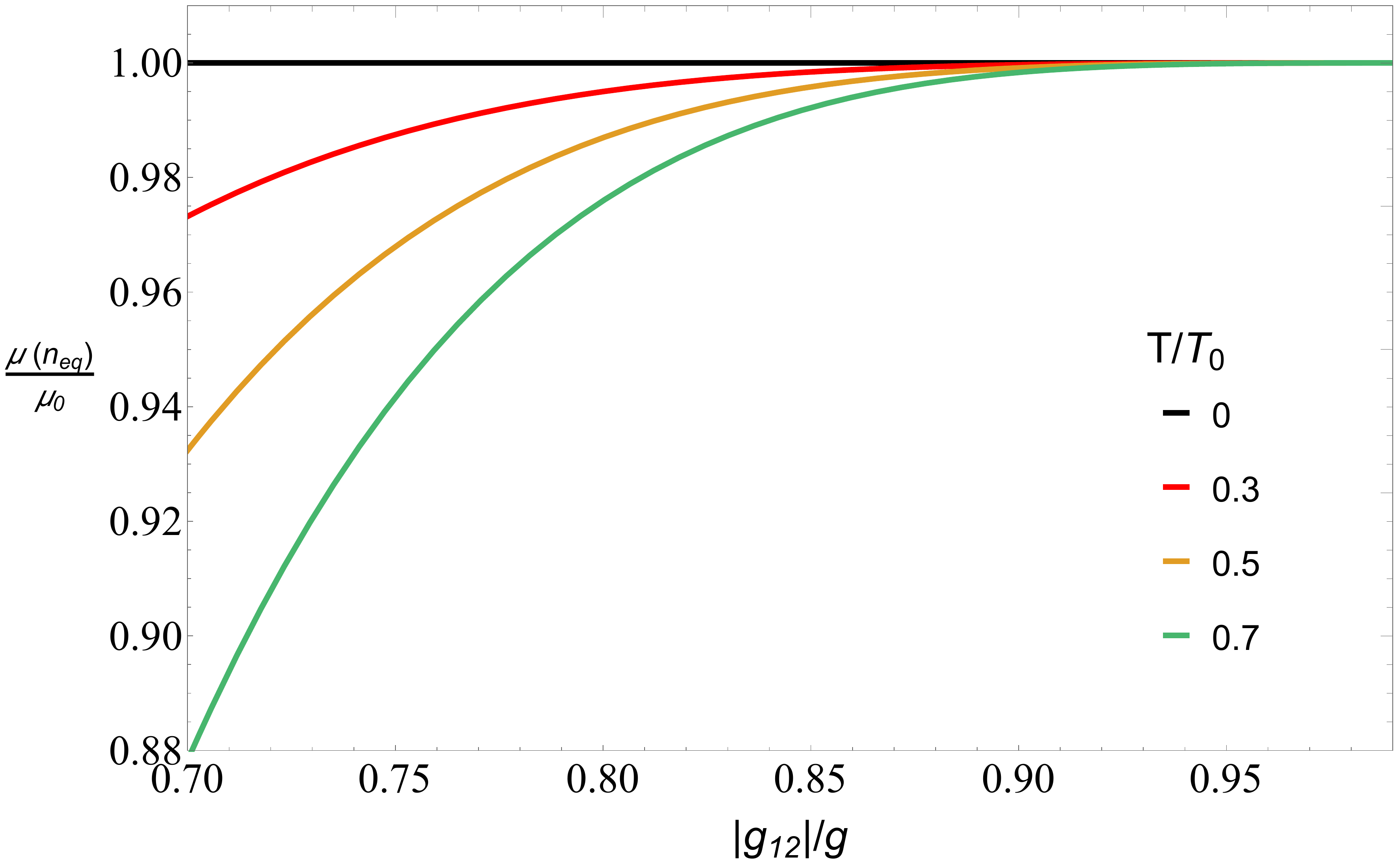}
\caption{Chemical potential as a function of $|g_{12}|/g$. Various colors correspond to different temperatures.~The curves are reported in an increasing order of the temperature from low (top) to high (bottom) values.~The chemical potential is reported in units of its zero-temperature value, Eq.~\eqref{Eq:mu0}, at $n = n_{\rm eq}$.
}
 \label{fig:mu}
 \end{center}
\end{figure}

\begin{figure}[t]
\begin{center}
\includegraphics[width=\columnwidth,angle=0,clip=]{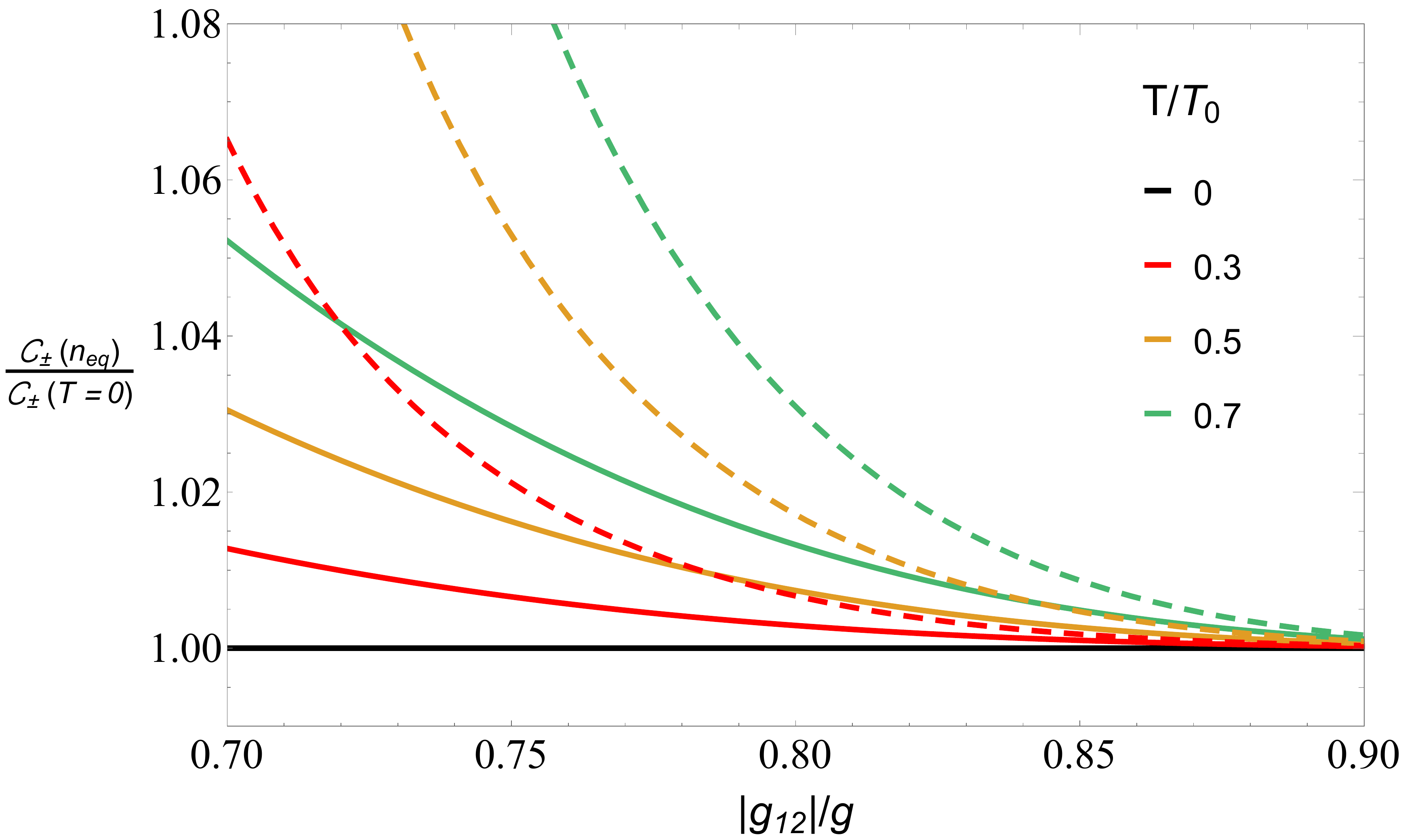}
\caption{Intra- ($\mathcal{C}_+$) (dashed) and inter- ($\mathcal{C}_-$) (solid) species Tan's contact densities as a function of $|g_{12}|/g$ for different values of temperature.~The curves are reported in an increasing order of the temperature from low (bottom) to high (top) values.~The contact densities are reported in units of their zero-temperature value, Eq.~\eqref{Eq:C0}, calculated at $n = n_{\rm eq}$.
}
 \label{fig:C}
 \end{center}
\end{figure}

\begin{figure}[t]
\begin{center}
\includegraphics[width=\columnwidth,angle=0,clip=]{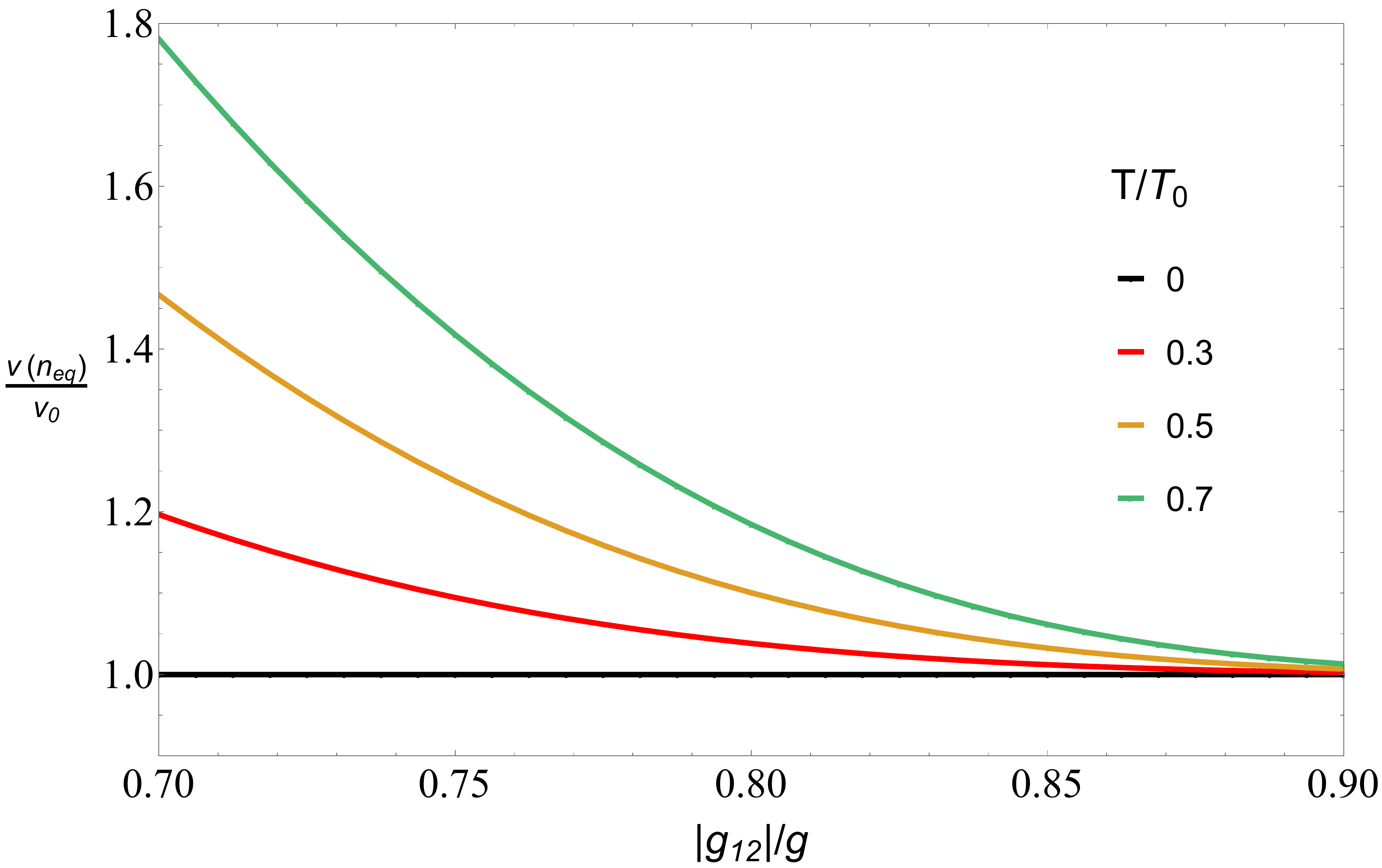}
\caption{Adiabatic sound velocity $v$, from Eq.~\eqref{Eq:mv2} as a function of $|g_{12}|/g$.~Various colors correspond to different temperatures.~The curves are reported in an increasing order of the temperature from low (bottom) to high (top) values.~The adiabatic sound velocity is reported in units of its zero-temperature value, Eq.~\eqref{Eq:v0}, at $n = n_{\rm eq}$.
}
\label{fig:v}
\end{center}
\end{figure}

\begin{figure}[t]
\begin{center}
\includegraphics[width=\columnwidth,angle=0,clip=]{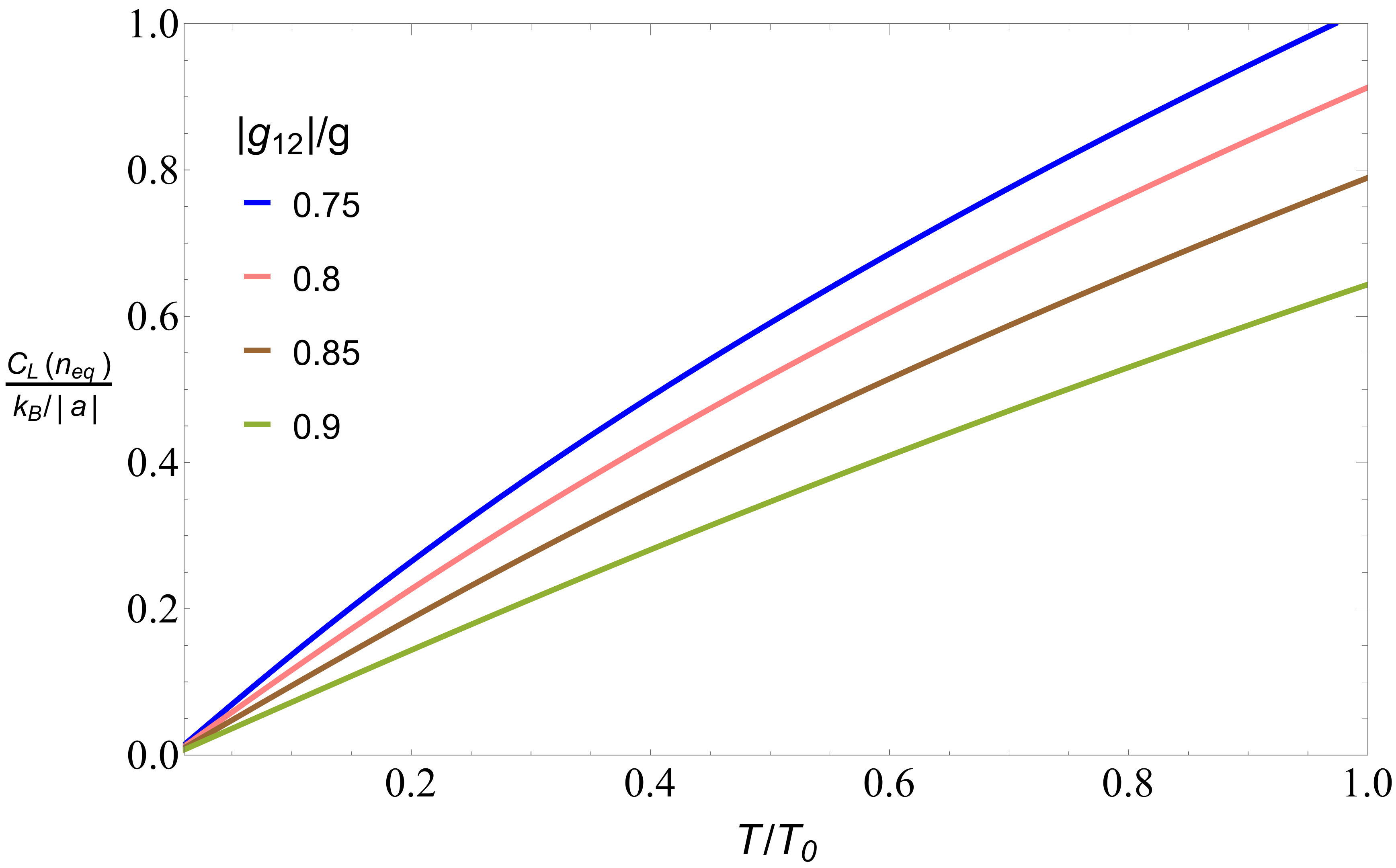}
\caption{
Specific heat density at constant length as a function of temperature, for different values of $|g_{12}|/g$.~The curves are reported in an increasing order of the ratio of the coupling constants from low (top) to high (bottom) values.
}
 \label{fig:Heat_neq}
 \end{center}
\end{figure}

After discussing the stability and the evaporation of the liquid, we proceed to the calculation of the main thermodynamic quantities from Eq.~\eqref{Eq:thermal A quasi-particles}. 
All results presented in this Section refer to systems at a density fixed by the equilibrium value at zero temperature $n_{\rm eq}$, Eq.~\eqref{Eq:neq}, and they are reported in Figs.~\ref{fig:mu}-\ref{fig:Heat_neq}. They are also valid then for saturated droplets.  

The thermal contribution to the chemical potential can be expressed as
\begin{equation}
\label{Eq:Delta_mu}
\Delta \mu =\frac{1}{2 n} \sum_\pm 
\int_{-\infty}^{+\infty} \frac{dp}{2 \pi \hbar}  \frac{p^2 c_\pm^2}{E_\pm(p)}   f\left( E_\pm\right) .
\end{equation}
Figure~\ref{fig:mu} shows the chemical potential which is always negative (even at zero temperature, Eq.~\eqref{Eq:mu0}), reflecting the bound nature of the liquid. 
Its absolute value decreases with temperature as $\Delta\mu$ is positive. This is in a striking contrast to that of a single-component 1D Bose gas at low temperature: $\Delta \mu$ is also positive
but, since $\mu >0$, the effect of thermal fluctuations is to increase $|\mu|$  \cite{Mora2003, DeRosi2017, DeRosi2019}.

The thermal parts of the intra-species ($\Delta \mathcal{C}_+$) and inter-species ($\Delta \mathcal{C}_-$) Tan's contact parameter densities are:
\begin{multline}
\label{Eq:DeltaC}
\Delta \mathcal{C}_\pm = \frac{m^3}{2 n \hbar^4} \left(c_+^2 \pm c_-^2  \right)^2 \\ \int_{-\infty}^{+\infty} \frac{dp}{2 \pi \hbar} p^2 \left[\frac{f\left(E_- \right)}{E_-(p)} \pm \frac{f\left(E_+   \right)}{E_+(p)}  \right] .
\end{multline}
We recall that for a weakly-interacting single-component 1D Bose gas there is an intrinsic relation between the thermal correction of the contact and of the chemical potential $\Delta \mathcal{C}\propto\Delta \mu$  \cite{DeRosi2019}. Due to the non-trivial coupling between density and spin modes, such a direct proportionality is absent for two-component mixtures.
In Fig.~\ref{fig:C}, we observe that the intra-species Tan's contact density (dashed lines), which is determined by the repulsive interaction $g > 0$, is more sensitive to the thermal effects than the inter-species one (solid lines), which instead emerges from the attractive interaction $g_{12} < 0$. 

The adiabatic sound velocity $v$ can be calculated from Eq.~\eqref{Eq:v} (see Appendix \ref{Sec:Inverse adiabatic compressibility}):
\begin{equation}
\label{Eq:mv2}
m v^2 = m v_0^2 + n \Delta \kappa_T^{-1} - n^2 \left( \frac{\partial \bar{s}}{\partial n} \right)_{T} 
\left( \frac{\partial T}{\partial n} \right)_{\bar{s}} 
\end{equation}
where the entropy per particle is 
\begin{equation}
\label{Eq:sbar}
\bar{s} = \frac{1}{nT} \left[ \sum_\pm \int_{-\infty}^{+\infty} \frac{dp}{2 \pi \hbar} E_\pm(p) f(E_\pm) - \Delta\mathcal{A} \right].
\end{equation} 
In Fig.~\ref{fig:v} we show that the sound speed of the liquid increases with temperature as it occurs in a single-component 1D Bose gas  \cite{DeRosi2017} and in classical systems. 
Moreover, $v$ is greater for smaller values of $|g_{12}|/g$, as it occurs at zero temperature  \cite{Parisi2019}. 

The specific heat density, Eq.~\eqref{Eq:specific heat}, is:
\begin{equation}
\label{Eq:heatT}
C_L = n T \left(  \frac{\partial \bar{s}}{\partial T}   \right)_n
\end{equation}
where the derivative of $\bar{s}$ is provided by Eq.~\eqref{Eq:DsbarT}. Our results are shown in Fig.~\ref{fig:Heat_neq}.

\section{Experimental Considerations}
\label{Sec:experiments}

The standard thermometry technique, which is based on switching off the trap and studying the expansion during time-of-flight, cannot be applied to quantum liquids  \cite{Bottcher2021}. 
One possible method is to introduce impurities and perform spectroscopic measurements \cite{Wenzel2018}, similarly to the experiments on liquid helium droplets  \cite{Toennies2004}, 
but this complicates the setup as it requires a simultaneous handling of three components. 
Another temperature probe is based on the measurements of the density fluctuations and compressibility which directly yield a theory-independent thermometer via the fluctuation-dissipation theorem  \cite{Hartke2020}. 
Indeed the temperature enters in the fluctuation-dissipation theorem, which relates the compressibility to the global number fluctuations of the system. This approach requires measurements of the total density sensitive to atomic shot noise.

Here we suggest that a simpler method for probing the temperature in quantum liquid experiments would be the direct non-destructive in-situ measurement of the thermodynamic properties. In trapped gases, a single measurement of the density profile provides access to the complete equation of state of the system  \cite{Ho2009}. The pressure, the chemical potential, the isothermal compressibility, the specific heat per particle at constant volume, the free energy, the energy and the entropy per particle as a function of temperature have been measured in-situ by using absorption imaging in 3D gases  \cite{Ho2009, Nascimbene2010b, Ku2012}. 

A similar experimental technique has been applied to 1D Bose gas to extract the chemical potential as a function of temperature and interaction strength  \cite{Salces-Carcoba2018}. In-situ thermometry through high-resolution absorption imaging measurements of the thermodynamics has been recently achieved in 1D fermionic balanced mixtures \cite{DeDaniloff2021}. Tan's contact can be measured with Bragg spectroscopy  \cite{Hoinka2013} and the sound velocity can be extracted by exciting the system locally and by observing the propagation speed of the density perturbation, especially in a quasi-1D geometry, by employing phase-contrast techniques  \cite{Andrews1997, Joseph2007}.

In alternative to the direct measurement of the temperature, which can be problematic in the very low-$T$ regime, one can instead combine the experimental estimates of several thermodynamic quantities \cite{Pitaevskii2016}. An example can be provided by the measurements of the Tan's contacts, the free energy and the density, so that the temperature of the system can then be extracted \textit{a posteriori} through the use of Eq.~\eqref{Eq:contact} and our Fig.~\ref{fig:C} as benchmark. A similar procedure involving the measurements of the density, the isothermal compressibility and the pressure has been successfully applied in Fermi \cite{Ku2012} and Bose \cite{Desbuquois2014} gases.

Differently from trapped gases, in-situ measurements on self-bound uniform liquids are not affected by density inhomogeneity effects. This promises high-precision measurements in ultracold liquids.

One-dimensional configurations can be achieved in current quantum liquid experiments  \cite{Cheiney2018}, and this leads to strongly reduced three-body losses \cite{Tolra2004,Lavoine2021}. As such both the thermal dynamical instability and the evaporation discussed in this paper could be directly and unambiguously observed. Our results also suggest that quantum liquids may be experimentally obtained from the gas state by direct cooling. 
Finally, the strong dependence of the dynamical instability critical temperature on $|g_{12}|/g$, which can be fine-tuned in experiments,   provides yet another appealing probe of temperature for quantum liquids. 

\section{Conclusions and Outlooks}
\label{Sec:conclusions}

In this paper, we investigated the low-temperature behavior of the 1D weakly-interacting liquids formed in two-component Bose mixtures.
We discussed in detail the dynamical instability and the evaporation which drive the thermal liquid-to-gas transition and we calculated the main thermodynamic quantities of the liquid. 
Our results are based on the Bogoliubov theory, which allows for a simple description of the system in terms of non-interacting bosonic excitations and the inclusion of the quantum fluctuations in the thermodynamics. We provided the phase diagram of the system in terms of the interaction strengths and the temperature. 
Such information can be of great help to realize quantum liquids by directly lowering the temperature of the gas. Consistently with Ref.~\cite{Ota2020}, the dominant contribution to the thermal effects is provided by the excitation of the soft density mode. 
We have also provided a thorough study of the thermodynamic quantities of the liquid at low temperature, which are of a large importance to the experiments: the chemical potential, the intra- and inter-species Tan's contact parameter densities, the adiabatic sound velocity and the specific heat density at constant length. 

Our theoretical predictions suggest novel important precise temperature probes for the experiments on quantum liquids, where time-of-flight expansion cannot be applied and density inhomogeneity effects are strongly reduced. The critical temperature for dynamical instability can be fine-tuned at will with Fano-Feshbach resonances  \cite{DErrico2007, Chin2010, Tanzi2018}.
Moreover, one can apply in-situ measurements of the thermodynamic quantities. 
Both methods can be employed to measure the temperature in quantum liquids with unprecedented precision and our results constitute a fundamental benchmark. 
We found that the dynamical instability and the evaporation take place at low and high liquid densities, respectively. 
Both thermal mechanisms can be observed with high precision and their onset is completely reversible by tuning the interaction strengths. 
Here we studied uniform liquids, but all our findings at zero pressure apply directly also to saturated droplets. 

In outlook, our results can stimulate further theoretical and experimental investigations aiming at the characterization of quantum degenerate phases in 1D weakly-interacting mixtures at finite temperature and the microscopic nature of new 1D liquids. Our finite-temperature analysis can be generalized to: i) unsaturated droplets  \cite{Petrov2016}, even in 1D optical lattices  \cite{Zhou2019, Morera2020, Morera20202}; 
ii) the inclusion of higher-order corrections in density and spin sound velocities  \cite{Ota2020}, in order to understand their effects on thermal quantum fluctuations; iii) dimerized liquids emerging from attractive atomic mixtures  \cite{Pricoupenko2018, Guijarro2018}. 
Also, the knowledge of thermodynamics is crucial for predicting the temperature dependence of the breathing modes \cite{DeRosi2015, DeRosi2016}. Theoretically, breathing modes of 1D liquids have been studied only at zero temperature  \cite{Astrakharchik2018, Parisi2020, Tylutki2020}.
So far, no collective modes have been observed yet due to the short experimental lifetime of liquids. Our predictions are also relevant for Rabi-coupled Bose-Bose mixtures, where low-$T$ quantum fluctuations drive the emergence of a droplet, which evaporates above a critical Rabi frequency  \cite{Cappellaro2017}.
Other interesting perspectives regard the investigation of the properties of impurities.
One can consider baths of very different nature, like: helium  \cite{Bardeen1967, Stienkemeier2006}, ultracold liquids  \cite{Wenzel2018, Bisset2021} and other 1D quantum liquids  \cite{Recati2005, Schecter2014}.
Thermal effects can be also investigated in the dynamical formation of liquids via evaporation  \cite{Ferioli2020} and in collisions between droplets  \cite{Astrakharchik2018, Ferioli2019}. 
Interesting perspectives of our results open in liquids with: i) different atomic species which have been realized even in quasi-1D geometry  \cite{DErrico2019, Burchianti2020}; and ii) different component densities or intra-species interaction strengths  \cite{Mithun2020}. 
Another exciting perspective is provided by the mixed bubbles in weakly-repulsive bosonic mixtures  \cite{Naidon2021}.

\begin{acknowledgments} 

The authors gratefully acknowledge 
D.~Rakshit 
for stimulating exchanges in the early stage of this research. 
They also thank J.~Boronat, D.~Petrov, C.~R.~Cabrera and  L.~A.~Pe\~{n}a~Ardila for insightful discussions. 

G.~D.~R.'s received funding from the European Union's Horizon 2020 research and innovation program under the Marie Sk\l odowska-Curie grant agreement {\it UltraLiquid} No. 797684.
All authors acknowledge financial support from the Spanish MINECO (FIS2017-84114-C2-1-P), and from the
Secretaria d'Universitats i Recerca del Departament d'Empresa i Coneixement de la Generalitat de Catalunya 
 within the ERDF Operational Program of Catalunya (project QuantumCat, Ref.~001-P-001644).
\end{acknowledgments}

\appendix

\section{Thermodynamic relations}
\label{Sec:thermodynamic relations}

In this Appendix, we provide details about the derivation of the thermodynamic relations presented in Eq.~\eqref{Eq:thermodynamic identities}.
Very general considerations on dimensional analysis  \cite{FetterBook, Barth2011, Patu2017} strongly constrain the functional form of the free energy density $\mathcal{A} = \mathcal{E} - T \mathcal{S}$:
\begin{equation}
\label{Eq:A general}
\mathcal{A}(T, a,a_{12}, n) \propto n^3 f\left(n a, n a_{12}, T/n^2 \right).\end{equation}
From Eq.~\eqref{Eq:A general}, one can deduce the scaling law
\begin{equation}
\label{Eq:A scaling law}
\mathcal{A}(\ell^2T,\ell^{-1} a, \ell^{-1} a_{12}, \ell n) = \ell^3 \mathcal{A}(T, a, a_{12}, n)
\end{equation}
where $\ell$ is an arbitrary, dimensionless parameter.
Taking the derivative of Eq.~\eqref{Eq:A scaling law} with respect to $\ell$ at $\ell = 1$ yields
\begin{multline}
\label{Eq:A Tan}
\Bigl[  2 T \left(\frac{\partial }{\partial T}\right)_{a, a_{12}, n} - a \left(\frac{\partial }{\partial a}\right)_{T, a_{12}, n} - a_{12} \left(\frac{\partial }{\partial a_{12}}\right)_{T, a, n} \\ + n \left(\frac{\partial}{\partial n}\right)_{T, a, a_{12}}   \Bigr] \mathcal{A} (T, a, a_{12}, n) = 3 \mathcal{A} (T, a, a_{12}, n) .
\end{multline}
From Eq.~\eqref{Eq:A Tan} and by using Eqs.~\eqref{Eq:mu}-\eqref{Eq:S}, and Eq.~\eqref{Eq:contact}, we find Eq.~\eqref{Eq:thermodynamic identities}.
The latter is a generalization to two-species mixtures of the single-component result found in Ref.~  \cite{DeRosi2019}.

\section{Adiabatic sound velocity}
\label{Sec:Inverse adiabatic compressibility}

In this Appendix we provide details on the derivation of the adiabatic sound velocity.
Using Eq.~\eqref{Eq:v}, one gets:
\begin{equation}
\label{Eq:mv2 Appendix}
m v^2 = \left( \frac{\partial P}{\partial n}  \right)_{\bar{s}}  = \left( \frac{\partial P}{\partial n}     \right)_T + \left( \frac{\partial P}{\partial T}     \right)_n 
\left( \frac{\partial T}{\partial n} \right)_{\bar{s}}  .
\end{equation}
Given a function of the form $h(x, y, z) = 0$, the partial derivatives of its variables can be related with the triple product rule:
\begin{equation}
\left(   \frac{\partial x}{\partial y}  \right)_z
\left(   \frac{\partial y}{\partial z}  \right)_x
\left(   \frac{\partial z}{\partial x}  \right)_y = -1 .
\end{equation}
Hence, one can rewrite:
\begin{equation}
\label{Eq:dTdn}
\left( \frac{\partial T}{\partial n} \right)_{\bar{s}} = - \left( \frac{\partial \bar{s}}{\partial n} \right)_T/\left( \frac{\partial \bar{s}}{\partial T} \right)_n,
\end{equation}
where 
\begin{equation}
\left( \frac{\partial \bar{s}}{\partial n} \right)_T = - \frac{\bar{s}}{n} + \frac{\beta}{2 n^2 T} \sum_\pm     \int_{-\infty}^{+\infty} \frac{dp}{2 \pi \hbar}  \frac{p^2 c_\pm^2}{E_\pm(p)} \frac{\partial }{\partial \beta}  f(E_\pm)   
\end{equation}
and 
\begin{equation}
\label{Eq:DsbarT}
\left( \frac{\partial \bar{s}}{\partial T} \right)_n = - \frac{\beta}{n T^2} \sum_\pm     \int_{-\infty}^{+\infty} \frac{dp}{2 \pi \hbar}  E_\pm(p)  \frac{\partial }{\partial \beta}  f(E_\pm).
\end{equation}
In the derivation we have used Eqs.~(\ref{Eq:cpm2}-\ref{Eq:BG spectrum}) and \eqref{Eq:sbar}.

The derivatives of the pressure are: 
\begin{equation}
\left( \frac{\partial P}{\partial n}     \right)_T  = m v_0^2 + n \Delta \kappa_T^{-1}
\end{equation}
and 
\begin{equation}
\label{Eq:dPdT}
\left( \frac{\partial P}{\partial T}     \right)_n = - n^2 \left( \frac{\partial \bar{s}}{\partial n} \right)_T
\end{equation}
where we have applied Eq.~\eqref{Eq:v0} and Eq.~\eqref{Eq:kT general spectrum}.

Inserting Eqs.~\eqref{Eq:dTdn}-\eqref{Eq:dPdT} in Eq.~\eqref{Eq:mv2 Appendix} finally gives Eq.~\eqref{Eq:mv2}.

\bibliography{Bibliography}

\begin{thebibliography}{100}%
\makeatletter
\providecommand \@ifxundefined [1]{%
 \@ifx{#1\undefined}
}%
\providecommand \@ifnum [1]{%
 \ifnum #1\expandafter \@firstoftwo
 \else \expandafter \@secondoftwo
 \fi
}%
\providecommand \@ifx [1]{%
 \ifx #1\expandafter \@firstoftwo
 \else \expandafter \@secondoftwo
 \fi
}%
\providecommand \natexlab [1]{#1}%
\providecommand \enquote  [1]{``#1''}%
\providecommand \bibnamefont  [1]{#1}%
\providecommand \bibfnamefont [1]{#1}%
\providecommand \citenamefont [1]{#1}%
\providecommand \href@noop [0]{\@secondoftwo}%
\providecommand \href [0]{\begingroup \@sanitize@url \@href}%
\providecommand \@href[1]{\@@startlink{#1}\@@href}%
\providecommand \@@href[1]{\endgroup#1\@@endlink}%
\providecommand \@sanitize@url [0]{\catcode `\\12\catcode `\$12\catcode
  `\&12\catcode `\#12\catcode `\^12\catcode `\_12\catcode `\%12\relax}%
\providecommand \@@startlink[1]{}%
\providecommand \@@endlink[0]{}%
\providecommand \url  [0]{\begingroup\@sanitize@url \@url }%
\providecommand \@url [1]{\endgroup\@href {#1}{\urlprefix }}%
\providecommand \urlprefix  [0]{URL }%
\providecommand \Eprint [0]{\href }%
\providecommand \doibase [0]{http://dx.doi.org/}%
\providecommand \selectlanguage [0]{\@gobble}%
\providecommand \bibinfo  [0]{\@secondoftwo}%
\providecommand \bibfield  [0]{\@secondoftwo}%
\providecommand \translation [1]{[#1]}%
\providecommand \BibitemOpen [0]{}%
\providecommand \bibitemStop [0]{}%
\providecommand \bibitemNoStop [0]{.\EOS\space}%
\providecommand \EOS [0]{\spacefactor3000\relax}%
\providecommand \BibitemShut  [1]{\csname bibitem#1\endcsname}%
\let\auto@bib@innerbib\@empty
\bibitem [{\citenamefont {Landau}\ and\ \citenamefont
  {Lifshitz}(2013{\natexlab{a}})}]{Landau2013fluid}%
  \BibitemOpen
  \bibfield  {author} {\bibinfo {author} {\bibfnamefont {L.}~\bibnamefont
  {Landau}}\ and\ \bibinfo {author} {\bibfnamefont {E.}~\bibnamefont
  {Lifshitz}},\ }\href@noop {} {\emph {\bibinfo {title} {Fluid Mechanics: V.
  6}}}\ (\bibinfo  {publisher} {Elsevier Science, Amsterdam},\ \bibinfo {year}
  {2013})\BibitemShut {NoStop}%
\bibitem [{\citenamefont {Boronat}(1998)}]{Boronat1998}%
  \BibitemOpen
  \bibfield  {author} {\bibinfo {author} {\bibfnamefont {J.}~\bibnamefont
  {Boronat}},\ }\bibfield  {title} {\bibinfo {title} {\emph {Diffusion Monte
  Carlo for excited states: Application to liquid helium}},\ }in\ \href@noop {}
  {\emph {\bibinfo {booktitle} {Microscopic Quantum Many-Body Theories and
  Their Applications}}},\ \bibinfo {editor} {edited by\ \bibinfo {editor}
  {\bibfnamefont {J.}~\bibnamefont {Navarro}}\ and\ \bibinfo {editor}
  {\bibfnamefont {A.}~\bibnamefont {Polls}}}\ (\bibinfo  {publisher} {Springer
  Berlin},\ \bibinfo {year} {1998})\ pp.\ \bibinfo {pages}
  {359--379}\BibitemShut {NoStop}%
\bibitem [{\citenamefont {Casulleras}\ and\ \citenamefont
  {Boronat}(2000)}]{Casulleras2000}%
  \BibitemOpen
  \bibfield  {author} {\bibinfo {author} {\bibfnamefont {J.}~\bibnamefont
  {Casulleras}}\ and\ \bibinfo {author} {\bibfnamefont {J.}~\bibnamefont
  {Boronat}},\ }\bibfield  {title} {\bibinfo {title} {\emph {Progress in Monte
  Carlo Calculations of Fermi Systems: Normal Liquid $^{3}He$}},\ }\href
  {\doibase 10.1103/PhysRevLett.84.3121} {\bibfield  {journal} {\bibinfo
  {journal} {Phys. Rev. Lett.}\ }\textbf {\bibinfo {volume} {84}},\ \bibinfo
  {pages} {3121} (\bibinfo {year} {2000})}\BibitemShut {NoStop}%
\bibitem [{\citenamefont {Boronat}\ and\ \citenamefont
  {Casulleras}(2001)}]{Boronat2001}%
  \BibitemOpen
  \bibfield  {author} {\bibinfo {author} {\bibfnamefont {J.}~\bibnamefont
  {Boronat}}\ and\ \bibinfo {author} {\bibfnamefont {J.}~\bibnamefont
  {Casulleras}},\ }\bibfield  {title} {\bibinfo {title} {\emph {New
  perspectives in the application of the Diffusion Monte Carlo method to the
  study of liquid $^3He$}},\ }\href {\doibase 10.1142/S0217979201006070}
  {\bibfield  {journal} {\bibinfo  {journal} {Int. J. Mod. Phys. B}\ }\textbf
  {\bibinfo {volume} {15}},\ \bibinfo {pages} {1591} (\bibinfo {year}
  {2001})}\BibitemShut {NoStop}%
\bibitem [{\citenamefont {Dalfovo}\ and\ \citenamefont
  {Stringari}(2001)}]{Dalfovo2001}%
  \BibitemOpen
  \bibfield  {author} {\bibinfo {author} {\bibfnamefont {F.}~\bibnamefont
  {Dalfovo}}\ and\ \bibinfo {author} {\bibfnamefont {S.}~\bibnamefont
  {Stringari}},\ }\bibfield  {title} {\bibinfo {title} {\emph {Helium
  nanodroplets and trapped Bose-Einstein condensates as prototypes of finite
  quantum fluids}},\ }\href {\doibase 10.1063/1.1424926} {\bibfield  {journal}
  {\bibinfo  {journal} {J. Chem. Phys.}\ }\textbf {\bibinfo {volume} {115}},\
  \bibinfo {pages} {10078} (\bibinfo {year} {2001})}\BibitemShut {NoStop}%
\bibitem [{\citenamefont {Toennies}\ \emph {et~al.}(2001)\citenamefont
  {Toennies}, \citenamefont {Vilesov},\ and\ \citenamefont
  {Whaley}}]{Toennies2001}%
  \BibitemOpen
  \bibfield  {author} {\bibinfo {author} {\bibfnamefont {P.}~\bibnamefont
  {Toennies}}, \bibinfo {author} {\bibfnamefont {A.}~\bibnamefont {Vilesov}}, \
  and\ \bibinfo {author} {\bibfnamefont {B.}~\bibnamefont {Whaley}},\
  }\bibfield  {title} {\bibinfo {title} {\emph {Superfluid Helium Droplets: An
  Ultracold Nanolaboratory}},\ }\href {\doibase 10.1063/1.1359707} {\bibfield
  {journal} {\bibinfo  {journal} {Phys. Today}\ }\textbf {\bibinfo {volume}
  {54}},\ \bibinfo {pages} {31} (\bibinfo {year} {2001})}\BibitemShut {NoStop}%
\bibitem [{\citenamefont {Volovik}(2003)}]{Volovik2003}%
  \BibitemOpen
  \bibfield  {author} {\bibinfo {author} {\bibfnamefont {G.~E.}\ \bibnamefont
  {Volovik}},\ }\href@noop {} {\emph {\bibinfo {title} {The Universe in a
  Helium Droplet}}},\ International Series of Monographs on Physics\ (\bibinfo
  {publisher} {Oxford University Press, Oxford},\ \bibinfo {year}
  {2003})\BibitemShut {NoStop}%
\bibitem [{\citenamefont {Diallo}\ \emph {et~al.}(2012)\citenamefont {Diallo},
  \citenamefont {Azuah}, \citenamefont {Abernathy}, \citenamefont {Rota},
  \citenamefont {Boronat},\ and\ \citenamefont {Glyde}}]{Diallo2012}%
  \BibitemOpen
  \bibfield  {author} {\bibinfo {author} {\bibfnamefont {S.~O.}\ \bibnamefont
  {Diallo}}, \bibinfo {author} {\bibfnamefont {R.~T.}\ \bibnamefont {Azuah}},
  \bibinfo {author} {\bibfnamefont {D.~L.}\ \bibnamefont {Abernathy}}, \bibinfo
  {author} {\bibfnamefont {R.}~\bibnamefont {Rota}}, \bibinfo {author}
  {\bibfnamefont {J.}~\bibnamefont {Boronat}}, \ and\ \bibinfo {author}
  {\bibfnamefont {H.~R.}\ \bibnamefont {Glyde}},\ }\bibfield  {title} {\bibinfo
  {title} {\emph {Bose-Einstein condensation in liquid ${}^{4}$He near the
  liquid-solid transition line}},\ }\href {\doibase 10.1103/PhysRevB.85.140505}
  {\bibfield  {journal} {\bibinfo  {journal} {Phys. Rev. B}\ }\textbf {\bibinfo
  {volume} {85}},\ \bibinfo {pages} {140505(R)} (\bibinfo {year}
  {2012})}\BibitemShut {NoStop}%
\bibitem [{\citenamefont {B\"{o}ttcher}\ \emph {et~al.}(2021)\citenamefont
  {B\"{o}ttcher}, \citenamefont {Schmidt}, \citenamefont {Hertkorn},
  \citenamefont {Ng}, \citenamefont {Graham}, \citenamefont {Guo},
  \citenamefont {Langen},\ and\ \citenamefont {Pfau}}]{Bottcher2021}%
  \BibitemOpen
  \bibfield  {author} {\bibinfo {author} {\bibfnamefont {F.}~\bibnamefont
  {B\"{o}ttcher}}, \bibinfo {author} {\bibfnamefont {J.-N.}\ \bibnamefont
  {Schmidt}}, \bibinfo {author} {\bibfnamefont {J.}~\bibnamefont {Hertkorn}},
  \bibinfo {author} {\bibfnamefont {K.~S.~H.}\ \bibnamefont {Ng}}, \bibinfo
  {author} {\bibfnamefont {S.~D.}\ \bibnamefont {Graham}}, \bibinfo {author}
  {\bibfnamefont {M.}~\bibnamefont {Guo}}, \bibinfo {author} {\bibfnamefont
  {T.}~\bibnamefont {Langen}}, \ and\ \bibinfo {author} {\bibfnamefont
  {T.}~\bibnamefont {Pfau}},\ }\bibfield  {title} {\bibinfo {title} {\emph {New
  states of matter with fine-tuned interactions: quantum droplets and dipolar
  supersolids}},\ }\href {\doibase 10.1088/1361-6633/abc9ab} {\bibfield
  {journal} {\bibinfo  {journal} {Rep. Prog. Phys.}\ }\textbf {\bibinfo
  {volume} {84}},\ \bibinfo {pages} {012403} (\bibinfo {year}
  {2021})}\BibitemShut {NoStop}%
\bibitem [{\citenamefont {Luo}\ \emph {et~al.}(2021)\citenamefont {Luo},
  \citenamefont {Pang}, \citenamefont {Liu}, \citenamefont {Li},\ and\
  \citenamefont {Malomed}}]{Luo2021}%
  \BibitemOpen
  \bibfield  {author} {\bibinfo {author} {\bibfnamefont {Z.-H.}\ \bibnamefont
  {Luo}}, \bibinfo {author} {\bibfnamefont {W.}~\bibnamefont {Pang}}, \bibinfo
  {author} {\bibfnamefont {B.}~\bibnamefont {Liu}}, \bibinfo {author}
  {\bibfnamefont {Y.-Y.}\ \bibnamefont {Li}}, \ and\ \bibinfo {author}
  {\bibfnamefont {B.~A.}\ \bibnamefont {Malomed}},\ }\bibfield  {title}
  {\bibinfo {title} {\emph {A new form of liquid matter: Quantum droplets}},\
  }\href {https://doi.org/10.1007/s11467-020-1020-2} {\bibfield  {journal}
  {\bibinfo  {journal} {Front. Phys.}\ }\textbf {\bibinfo {volume} {16}},\
  \bibinfo {pages} {32201} (\bibinfo {year} {2021})}\BibitemShut {NoStop}%
\bibitem [{\citenamefont {Kadau}\ \emph {et~al.}(2016)\citenamefont {Kadau},
  \citenamefont {Schmitt}, \citenamefont {Wenzel}, \citenamefont {Wink},
  \citenamefont {Maier}, \citenamefont {Ferrier-Barbut},\ and\ \citenamefont
  {Pfau}}]{Kadau2016}%
  \BibitemOpen
  \bibfield  {author} {\bibinfo {author} {\bibfnamefont {H.}~\bibnamefont
  {Kadau}}, \bibinfo {author} {\bibfnamefont {M.}~\bibnamefont {Schmitt}},
  \bibinfo {author} {\bibfnamefont {M.}~\bibnamefont {Wenzel}}, \bibinfo
  {author} {\bibfnamefont {C.}~\bibnamefont {Wink}}, \bibinfo {author}
  {\bibfnamefont {T.}~\bibnamefont {Maier}}, \bibinfo {author} {\bibfnamefont
  {I.}~\bibnamefont {Ferrier-Barbut}}, \ and\ \bibinfo {author} {\bibfnamefont
  {T.}~\bibnamefont {Pfau}},\ }\bibfield  {title} {\bibinfo {title} {\emph
  {Observing the Rosensweig instability of a quantum ferrofluid}},\ }\href
  {\doibase https://doi.org/10.1038/nature16485} {\bibfield  {journal}
  {\bibinfo  {journal} {Nature}\ }\textbf {\bibinfo {volume} {530}},\ \bibinfo
  {pages} {194} (\bibinfo {year} {2016})}\BibitemShut {NoStop}%
\bibitem [{\citenamefont {Ferrier-Barbut}\ \emph
  {et~al.}(2016{\natexlab{a}})\citenamefont {Ferrier-Barbut}, \citenamefont
  {Kadau}, \citenamefont {Schmitt}, \citenamefont {Wenzel},\ and\ \citenamefont
  {Pfau}}]{Ferrier-Barbut2016}%
  \BibitemOpen
  \bibfield  {author} {\bibinfo {author} {\bibfnamefont {I.}~\bibnamefont
  {Ferrier-Barbut}}, \bibinfo {author} {\bibfnamefont {H.}~\bibnamefont
  {Kadau}}, \bibinfo {author} {\bibfnamefont {M.}~\bibnamefont {Schmitt}},
  \bibinfo {author} {\bibfnamefont {M.}~\bibnamefont {Wenzel}}, \ and\ \bibinfo
  {author} {\bibfnamefont {T.}~\bibnamefont {Pfau}},\ }\bibfield  {title}
  {\bibinfo {title} {\emph {Observation of Quantum Droplets in a Strongly
  Dipolar Bose Gas}},\ }\href {\doibase 10.1103/PhysRevLett.116.215301}
  {\bibfield  {journal} {\bibinfo  {journal} {Phys. Rev. Lett.}\ }\textbf
  {\bibinfo {volume} {116}},\ \bibinfo {pages} {215301} (\bibinfo {year}
  {2016}{\natexlab{a}})}\BibitemShut {NoStop}%
\bibitem [{\citenamefont {W\"achtler}\ and\ \citenamefont
  {Santos}(2016)}]{Wachtler2016}%
  \BibitemOpen
  \bibfield  {author} {\bibinfo {author} {\bibfnamefont {F.}~\bibnamefont
  {W\"achtler}}\ and\ \bibinfo {author} {\bibfnamefont {L.}~\bibnamefont
  {Santos}},\ }\bibfield  {title} {\bibinfo {title} {\emph {Quantum filaments
  in dipolar Bose-Einstein condensates}},\ }\href {\doibase
  10.1103/PhysRevA.93.061603} {\bibfield  {journal} {\bibinfo  {journal} {Phys.
  Rev. A}\ }\textbf {\bibinfo {volume} {93}},\ \bibinfo {pages} {061603(R)}
  (\bibinfo {year} {2016})}\BibitemShut {NoStop}%
\bibitem [{\citenamefont {Saito}(2016)}]{Saito2016}%
  \BibitemOpen
  \bibfield  {author} {\bibinfo {author} {\bibfnamefont {H.}~\bibnamefont
  {Saito}},\ }\bibfield  {title} {\bibinfo {title} {\emph {Path-Integral Monte
  Carlo Study on a Droplet of a Dipolar Bose-Einstein Condensate Stabilized by
  Quantum Fluctuation}},\ }\href {https://doi.org/10.7566/JPSJ.85.053001}
  {\bibfield  {journal} {\bibinfo  {journal} {J. Phys. Soc. Jpn.}\ }\textbf
  {\bibinfo {volume} {85}},\ \bibinfo {pages} {053001} (\bibinfo {year}
  {2016})}\BibitemShut {NoStop}%
\bibitem [{\citenamefont {Schmitt}\ \emph {et~al.}(2016)\citenamefont
  {Schmitt}, \citenamefont {Wenzel}, \citenamefont {B\"{o}ttcher},
  \citenamefont {Ferrier-Barbut},\ and\ \citenamefont {Pfau}}]{Schmitt2016}%
  \BibitemOpen
  \bibfield  {author} {\bibinfo {author} {\bibfnamefont {M.}~\bibnamefont
  {Schmitt}}, \bibinfo {author} {\bibfnamefont {M.}~\bibnamefont {Wenzel}},
  \bibinfo {author} {\bibfnamefont {F.}~\bibnamefont {B\"{o}ttcher}}, \bibinfo
  {author} {\bibfnamefont {I.}~\bibnamefont {Ferrier-Barbut}}, \ and\ \bibinfo
  {author} {\bibfnamefont {T.}~\bibnamefont {Pfau}},\ }\bibfield  {title}
  {\bibinfo {title} {\emph {Self-bound droplets of a dilute magnetic quantum
  liquid}},\ }\href {\doibase https://doi.org/10.1038/nature20126} {\bibfield
  {journal} {\bibinfo  {journal} {Nature}\ }\textbf {\bibinfo {volume} {539}},\
  \bibinfo {pages} {259} (\bibinfo {year} {2016})}\BibitemShut {NoStop}%
\bibitem [{\citenamefont {Ferrier-Barbut}\ \emph
  {et~al.}(2016{\natexlab{b}})\citenamefont {Ferrier-Barbut}, \citenamefont
  {Schmitt}, \citenamefont {Wenzel}, \citenamefont {Kadau},\ and\ \citenamefont
  {Pfau}}]{FerrierBarbut2016}%
  \BibitemOpen
  \bibfield  {author} {\bibinfo {author} {\bibfnamefont {I.}~\bibnamefont
  {Ferrier-Barbut}}, \bibinfo {author} {\bibfnamefont {M.}~\bibnamefont
  {Schmitt}}, \bibinfo {author} {\bibfnamefont {M.}~\bibnamefont {Wenzel}},
  \bibinfo {author} {\bibfnamefont {H.}~\bibnamefont {Kadau}}, \ and\ \bibinfo
  {author} {\bibfnamefont {T.}~\bibnamefont {Pfau}},\ }\bibfield  {title}
  {\bibinfo {title} {\emph {Liquid quantum droplets of ultracold magnetic
  atoms}},\ }\href {https://doi.org/10.1088%2F0953-4075%2F49%2F21%2F214004}
  {\bibfield  {journal} {\bibinfo  {journal} {J. Phys. B: At. Mol. Opt. Phys.}\
  }\textbf {\bibinfo {volume} {49}},\ \bibinfo {pages} {214004} (\bibinfo
  {year} {2016}{\natexlab{b}})}\BibitemShut {NoStop}%
\bibitem [{\citenamefont {Chomaz}\ \emph {et~al.}(2016)\citenamefont {Chomaz},
  \citenamefont {Baier}, \citenamefont {Petter}, \citenamefont {Mark},
  \citenamefont {W\"achtler}, \citenamefont {Santos},\ and\ \citenamefont
  {Ferlaino}}]{Chomaz2016}%
  \BibitemOpen
  \bibfield  {author} {\bibinfo {author} {\bibfnamefont {L.}~\bibnamefont
  {Chomaz}}, \bibinfo {author} {\bibfnamefont {S.}~\bibnamefont {Baier}},
  \bibinfo {author} {\bibfnamefont {D.}~\bibnamefont {Petter}}, \bibinfo
  {author} {\bibfnamefont {M.~J.}\ \bibnamefont {Mark}}, \bibinfo {author}
  {\bibfnamefont {F.}~\bibnamefont {W\"achtler}}, \bibinfo {author}
  {\bibfnamefont {L.}~\bibnamefont {Santos}}, \ and\ \bibinfo {author}
  {\bibfnamefont {F.}~\bibnamefont {Ferlaino}},\ }\bibfield  {title} {\bibinfo
  {title} {\emph {Quantum-Fluctuation-Driven Crossover from a Dilute
  Bose-Einstein Condensate to a Macrodroplet in a Dipolar Quantum Fluid}},\
  }\href {\doibase 10.1103/PhysRevX.6.041039} {\bibfield  {journal} {\bibinfo
  {journal} {Phys. Rev. X}\ }\textbf {\bibinfo {volume} {6}},\ \bibinfo {pages}
  {041039} (\bibinfo {year} {2016})}\BibitemShut {NoStop}%
\bibitem [{\citenamefont {B\"ottcher}\ \emph {et~al.}(2019)\citenamefont
  {B\"ottcher}, \citenamefont {Wenzel}, \citenamefont {Schmidt}, \citenamefont
  {Guo}, \citenamefont {Langen}, \citenamefont {Ferrier-Barbut}, \citenamefont
  {Pfau}, \citenamefont {Bomb\'{\i}n}, \citenamefont {S\'anchez-Baena},
  \citenamefont {Boronat},\ and\ \citenamefont {Mazzanti}}]{Bottcher2019}%
  \BibitemOpen
  \bibfield  {author} {\bibinfo {author} {\bibfnamefont {F.}~\bibnamefont
  {B\"ottcher}}, \bibinfo {author} {\bibfnamefont {M.}~\bibnamefont {Wenzel}},
  \bibinfo {author} {\bibfnamefont {J.-N.}\ \bibnamefont {Schmidt}}, \bibinfo
  {author} {\bibfnamefont {M.}~\bibnamefont {Guo}}, \bibinfo {author}
  {\bibfnamefont {T.}~\bibnamefont {Langen}}, \bibinfo {author} {\bibfnamefont
  {I.}~\bibnamefont {Ferrier-Barbut}}, \bibinfo {author} {\bibfnamefont
  {T.}~\bibnamefont {Pfau}}, \bibinfo {author} {\bibfnamefont {R.}~\bibnamefont
  {Bomb\'{\i}n}}, \bibinfo {author} {\bibfnamefont {J.}~\bibnamefont
  {S\'anchez-Baena}}, \bibinfo {author} {\bibfnamefont {J.}~\bibnamefont
  {Boronat}}, \ and\ \bibinfo {author} {\bibfnamefont {F.}~\bibnamefont
  {Mazzanti}},\ }\bibfield  {title} {\bibinfo {title} {\emph {Dilute dipolar
  quantum droplets beyond the extended Gross-Pitaevskii equation}},\ }\href
  {\doibase 10.1103/PhysRevResearch.1.033088} {\bibfield  {journal} {\bibinfo
  {journal} {Phys. Rev. Res.}\ }\textbf {\bibinfo {volume} {1}},\ \bibinfo
  {pages} {033088} (\bibinfo {year} {2019})}\BibitemShut {NoStop}%
\bibitem [{\citenamefont {Petrov}(2015)}]{Petrov2015}%
  \BibitemOpen
  \bibfield  {author} {\bibinfo {author} {\bibfnamefont {D.~S.}\ \bibnamefont
  {Petrov}},\ }\bibfield  {title} {\bibinfo {title} {\emph {Quantum Mechanical
  Stabilization of a Collapsing Bose-Bose Mixture}},\ }\href {\doibase
  10.1103/PhysRevLett.115.155302} {\bibfield  {journal} {\bibinfo  {journal}
  {Phys. Rev. Lett.}\ }\textbf {\bibinfo {volume} {115}},\ \bibinfo {pages}
  {155302} (\bibinfo {year} {2015})}\BibitemShut {NoStop}%
\bibitem [{\citenamefont {Cabrera}\ \emph {et~al.}(2018)\citenamefont
  {Cabrera}, \citenamefont {Tanzi}, \citenamefont {Sanz}, \citenamefont
  {Naylor}, \citenamefont {Thomas}, \citenamefont {Cheiney},\ and\
  \citenamefont {Tarruell}}]{Cabrera2018}%
  \BibitemOpen
  \bibfield  {author} {\bibinfo {author} {\bibfnamefont {C.~R.}\ \bibnamefont
  {Cabrera}}, \bibinfo {author} {\bibfnamefont {L.}~\bibnamefont {Tanzi}},
  \bibinfo {author} {\bibfnamefont {J.}~\bibnamefont {Sanz}}, \bibinfo {author}
  {\bibfnamefont {B.}~\bibnamefont {Naylor}}, \bibinfo {author} {\bibfnamefont
  {P.}~\bibnamefont {Thomas}}, \bibinfo {author} {\bibfnamefont
  {P.}~\bibnamefont {Cheiney}}, \ and\ \bibinfo {author} {\bibfnamefont
  {L.}~\bibnamefont {Tarruell}},\ }\bibfield  {title} {\bibinfo {title} {\emph
  {Quantum liquid droplets in a mixture of Bose-Einstein condensates}},\ }\href
  {\doibase 10.1126/science.aao5686} {\bibfield  {journal} {\bibinfo  {journal}
  {Science}\ }\textbf {\bibinfo {volume} {359}},\ \bibinfo {pages} {301}
  (\bibinfo {year} {2018})}\BibitemShut {NoStop}%
\bibitem [{\citenamefont {Semeghini}\ \emph {et~al.}(2018)\citenamefont
  {Semeghini}, \citenamefont {Ferioli}, \citenamefont {Masi}, \citenamefont
  {Mazzinghi}, \citenamefont {Wolswijk}, \citenamefont {Minardi}, \citenamefont
  {Modugno}, \citenamefont {Modugno}, \citenamefont {Inguscio},\ and\
  \citenamefont {Fattori}}]{Semeghini2018}%
  \BibitemOpen
  \bibfield  {author} {\bibinfo {author} {\bibfnamefont {G.}~\bibnamefont
  {Semeghini}}, \bibinfo {author} {\bibfnamefont {G.}~\bibnamefont {Ferioli}},
  \bibinfo {author} {\bibfnamefont {L.}~\bibnamefont {Masi}}, \bibinfo {author}
  {\bibfnamefont {C.}~\bibnamefont {Mazzinghi}}, \bibinfo {author}
  {\bibfnamefont {L.}~\bibnamefont {Wolswijk}}, \bibinfo {author}
  {\bibfnamefont {F.}~\bibnamefont {Minardi}}, \bibinfo {author} {\bibfnamefont
  {M.}~\bibnamefont {Modugno}}, \bibinfo {author} {\bibfnamefont
  {G.}~\bibnamefont {Modugno}}, \bibinfo {author} {\bibfnamefont
  {M.}~\bibnamefont {Inguscio}}, \ and\ \bibinfo {author} {\bibfnamefont
  {M.}~\bibnamefont {Fattori}},\ }\bibfield  {title} {\bibinfo {title} {\emph
  {Self-Bound Quantum Droplets of Atomic Mixtures in Free Space}},\ }\href
  {\doibase 10.1103/PhysRevLett.120.235301} {\bibfield  {journal} {\bibinfo
  {journal} {Phys. Rev. Lett.}\ }\textbf {\bibinfo {volume} {120}},\ \bibinfo
  {pages} {235301} (\bibinfo {year} {2018})}\BibitemShut {NoStop}%
\bibitem [{\citenamefont {Cikojevi\ifmmode~\acute{c}\else \'{c}\fi{}}\ \emph
  {et~al.}(2018)\citenamefont {Cikojevi\ifmmode~\acute{c}\else \'{c}\fi{}},
  \citenamefont {D\ifmmode~\check{z}\else \v{z}\fi{}elalija}, \citenamefont
  {Stipanovi\ifmmode~\acute{c}\else \'{c}\fi{}}, \citenamefont {Vranje\ifmmode
  \check{s}\else \v{s}\fi{} Marki\ifmmode~\acute{c}\else \'{c}\fi{}},\ and\
  \citenamefont {Boronat}}]{Cikojevic2017}%
  \BibitemOpen
  \bibfield  {author} {\bibinfo {author} {\bibfnamefont {V.}~\bibnamefont
  {Cikojevi\ifmmode~\acute{c}\else \'{c}\fi{}}}, \bibinfo {author}
  {\bibfnamefont {K.}~\bibnamefont {D\ifmmode~\check{z}\else
  \v{z}\fi{}elalija}}, \bibinfo {author} {\bibfnamefont {P.}~\bibnamefont
  {Stipanovi\ifmmode~\acute{c}\else \'{c}\fi{}}}, \bibinfo {author}
  {\bibfnamefont {L.}~\bibnamefont {Vranje\ifmmode \check{s}\else \v{s}\fi{}
  Marki\ifmmode~\acute{c}\else \'{c}\fi{}}}, \ and\ \bibinfo {author}
  {\bibfnamefont {J.}~\bibnamefont {Boronat}},\ }\bibfield  {title} {\bibinfo
  {title} {\emph {Ultradilute quantum liquid drops}},\ }\href {\doibase
  10.1103/PhysRevB.97.140502} {\bibfield  {journal} {\bibinfo  {journal} {Phys.
  Rev. B}\ }\textbf {\bibinfo {volume} {97}},\ \bibinfo {pages} {140502(R)}
  (\bibinfo {year} {2018})}\BibitemShut {NoStop}%
\bibitem [{\citenamefont {Ancilotto}\ \emph {et~al.}(2018)\citenamefont
  {Ancilotto}, \citenamefont {Barranco}, \citenamefont {Guilleumas},\ and\
  \citenamefont {Pi}}]{Ancilotto2018}%
  \BibitemOpen
  \bibfield  {author} {\bibinfo {author} {\bibfnamefont {F.}~\bibnamefont
  {Ancilotto}}, \bibinfo {author} {\bibfnamefont {M.}~\bibnamefont {Barranco}},
  \bibinfo {author} {\bibfnamefont {M.}~\bibnamefont {Guilleumas}}, \ and\
  \bibinfo {author} {\bibfnamefont {M.}~\bibnamefont {Pi}},\ }\bibfield
  {title} {\bibinfo {title} {\emph {Self-bound ultradilute Bose mixtures within
  local density approximation}},\ }\href {\doibase 10.1103/PhysRevA.98.053623}
  {\bibfield  {journal} {\bibinfo  {journal} {Phys. Rev. A}\ }\textbf {\bibinfo
  {volume} {98}},\ \bibinfo {pages} {053623} (\bibinfo {year}
  {2018})}\BibitemShut {NoStop}%
\bibitem [{\citenamefont {Petrov}\ and\ \citenamefont
  {Astrakharchik}(2016)}]{Petrov2016}%
  \BibitemOpen
  \bibfield  {author} {\bibinfo {author} {\bibfnamefont {D.~S.}\ \bibnamefont
  {Petrov}}\ and\ \bibinfo {author} {\bibfnamefont {G.~E.}\ \bibnamefont
  {Astrakharchik}},\ }\bibfield  {title} {\bibinfo {title} {\emph {Ultradilute
  Low-Dimensional Liquids}},\ }\href {\doibase 10.1103/PhysRevLett.117.100401}
  {\bibfield  {journal} {\bibinfo  {journal} {Phys. Rev. Lett.}\ }\textbf
  {\bibinfo {volume} {117}},\ \bibinfo {pages} {100401} (\bibinfo {year}
  {2016})}\BibitemShut {NoStop}%
\bibitem [{\citenamefont {Parisi}\ \emph {et~al.}(2019)\citenamefont {Parisi},
  \citenamefont {Astrakharchik},\ and\ \citenamefont {Giorgini}}]{Parisi2019}%
  \BibitemOpen
  \bibfield  {author} {\bibinfo {author} {\bibfnamefont {L.}~\bibnamefont
  {Parisi}}, \bibinfo {author} {\bibfnamefont {G.~E.}\ \bibnamefont
  {Astrakharchik}}, \ and\ \bibinfo {author} {\bibfnamefont {S.}~\bibnamefont
  {Giorgini}},\ }\bibfield  {title} {\bibinfo {title} {\emph {Liquid State of
  One-Dimensional Bose Mixtures: A Quantum Monte Carlo Study}},\ }\href
  {\doibase 10.1103/PhysRevLett.122.105302} {\bibfield  {journal} {\bibinfo
  {journal} {Phys. Rev. Lett.}\ }\textbf {\bibinfo {volume} {122}},\ \bibinfo
  {pages} {105302} (\bibinfo {year} {2019})}\BibitemShut {NoStop}%
\bibitem [{\citenamefont {Hu}\ \emph {et~al.}(2020)\citenamefont {Hu},
  \citenamefont {Wang},\ and\ \citenamefont {Liu}}]{Hu2020}%
  \BibitemOpen
  \bibfield  {author} {\bibinfo {author} {\bibfnamefont {H.}~\bibnamefont
  {Hu}}, \bibinfo {author} {\bibfnamefont {J.}~\bibnamefont {Wang}}, \ and\
  \bibinfo {author} {\bibfnamefont {X.-J.}\ \bibnamefont {Liu}},\ }\bibfield
  {title} {\bibinfo {title} {\emph {Microscopic pairing theory of a binary Bose
  mixture with interspecies attractions: Bosonic BEC-BCS crossover and
  ultradilute low-dimensional quantum droplets}},\ }\href {\doibase
  10.1103/PhysRevA.102.043301} {\bibfield  {journal} {\bibinfo  {journal}
  {Phys. Rev. A}\ }\textbf {\bibinfo {volume} {102}},\ \bibinfo {pages}
  {043301} (\bibinfo {year} {2020})}\BibitemShut {NoStop}%
\bibitem [{\citenamefont {Laburthe~Tolra}\ \emph {et~al.}(2004)\citenamefont
  {Laburthe~Tolra}, \citenamefont {O'Hara}, \citenamefont {Huckans},
  \citenamefont {Phillips}, \citenamefont {Rolston},\ and\ \citenamefont
  {Porto}}]{Tolra2004}%
  \BibitemOpen
  \bibfield  {author} {\bibinfo {author} {\bibfnamefont {B.}~\bibnamefont
  {Laburthe~Tolra}}, \bibinfo {author} {\bibfnamefont {K.~M.}\ \bibnamefont
  {O'Hara}}, \bibinfo {author} {\bibfnamefont {J.~H.}\ \bibnamefont {Huckans}},
  \bibinfo {author} {\bibfnamefont {W.~D.}\ \bibnamefont {Phillips}}, \bibinfo
  {author} {\bibfnamefont {S.~L.}\ \bibnamefont {Rolston}}, \ and\ \bibinfo
  {author} {\bibfnamefont {J.~V.}\ \bibnamefont {Porto}},\ }\bibfield  {title}
  {\bibinfo {title} {\emph {Observation of Reduced Three-Body Recombination in
  a Correlated 1D Degenerate Bose Gas}},\ }\href {\doibase
  10.1103/PhysRevLett.92.190401} {\bibfield  {journal} {\bibinfo  {journal}
  {Phys. Rev. Lett.}\ }\textbf {\bibinfo {volume} {92}},\ \bibinfo {pages}
  {190401} (\bibinfo {year} {2004})}\BibitemShut {NoStop}%
\bibitem [{\citenamefont {Lavoine}\ and\ \citenamefont
  {Bourdel}(2021)}]{Lavoine2021}%
  \BibitemOpen
  \bibfield  {author} {\bibinfo {author} {\bibfnamefont {L.}~\bibnamefont
  {Lavoine}}\ and\ \bibinfo {author} {\bibfnamefont {T.}~\bibnamefont
  {Bourdel}},\ }\bibfield  {title} {\bibinfo {title} {\emph {Beyond-mean-field
  crossover from one dimension to three dimensions in quantum droplets of
  binary mixtures}},\ }\href {\doibase 10.1103/PhysRevA.103.033312} {\bibfield
  {journal} {\bibinfo  {journal} {Phys. Rev. A}\ }\textbf {\bibinfo {volume}
  {103}},\ \bibinfo {pages} {033312} (\bibinfo {year} {2021})}\BibitemShut
  {NoStop}%
\bibitem [{\citenamefont {Cheiney}\ \emph {et~al.}(2018)\citenamefont
  {Cheiney}, \citenamefont {Cabrera}, \citenamefont {Sanz}, \citenamefont
  {Naylor}, \citenamefont {Tanzi},\ and\ \citenamefont
  {Tarruell}}]{Cheiney2018}%
  \BibitemOpen
  \bibfield  {author} {\bibinfo {author} {\bibfnamefont {P.}~\bibnamefont
  {Cheiney}}, \bibinfo {author} {\bibfnamefont {C.~R.}\ \bibnamefont
  {Cabrera}}, \bibinfo {author} {\bibfnamefont {J.}~\bibnamefont {Sanz}},
  \bibinfo {author} {\bibfnamefont {B.}~\bibnamefont {Naylor}}, \bibinfo
  {author} {\bibfnamefont {L.}~\bibnamefont {Tanzi}}, \ and\ \bibinfo {author}
  {\bibfnamefont {L.}~\bibnamefont {Tarruell}},\ }\bibfield  {title} {\bibinfo
  {title} {\emph {Bright Soliton to Quantum Droplet Transition in a Mixture of
  Bose-Einstein Condensates}},\ }\href {\doibase
  10.1103/PhysRevLett.120.135301} {\bibfield  {journal} {\bibinfo  {journal}
  {Phys. Rev. Lett.}\ }\textbf {\bibinfo {volume} {120}},\ \bibinfo {pages}
  {135301} (\bibinfo {year} {2018})}\BibitemShut {NoStop}%
\bibitem [{\citenamefont {Astrakharchik}\ and\ \citenamefont
  {Malomed}(2018)}]{Astrakharchik2018}%
  \BibitemOpen
  \bibfield  {author} {\bibinfo {author} {\bibfnamefont {G.~E.}\ \bibnamefont
  {Astrakharchik}}\ and\ \bibinfo {author} {\bibfnamefont {B.~A.}\ \bibnamefont
  {Malomed}},\ }\bibfield  {title} {\bibinfo {title} {\emph {Dynamics of
  one-dimensional quantum droplets}},\ }\href {\doibase
  10.1103/PhysRevA.98.013631} {\bibfield  {journal} {\bibinfo  {journal} {Phys.
  Rev. A}\ }\textbf {\bibinfo {volume} {98}},\ \bibinfo {pages} {013631}
  (\bibinfo {year} {2018})}\BibitemShut {NoStop}%
\bibitem [{\citenamefont {Parisi}\ and\ \citenamefont
  {Giorgini}(2020)}]{Parisi2020}%
  \BibitemOpen
  \bibfield  {author} {\bibinfo {author} {\bibfnamefont {L.}~\bibnamefont
  {Parisi}}\ and\ \bibinfo {author} {\bibfnamefont {S.}~\bibnamefont
  {Giorgini}},\ }\bibfield  {title} {\bibinfo {title} {\emph {Quantum droplets
  in one-dimensional Bose mixtures: A quantum Monte Carlo study}},\ }\href
  {\doibase 10.1103/PhysRevA.102.023318} {\bibfield  {journal} {\bibinfo
  {journal} {Phys. Rev. A}\ }\textbf {\bibinfo {volume} {102}},\ \bibinfo
  {pages} {023318} (\bibinfo {year} {2020})}\BibitemShut {NoStop}%
\bibitem [{\citenamefont {Tylutki}\ \emph {et~al.}(2020)\citenamefont
  {Tylutki}, \citenamefont {Astrakharchik}, \citenamefont {Malomed},\ and\
  \citenamefont {Petrov}}]{Tylutki2020}%
  \BibitemOpen
  \bibfield  {author} {\bibinfo {author} {\bibfnamefont {M.}~\bibnamefont
  {Tylutki}}, \bibinfo {author} {\bibfnamefont {G.~E.}\ \bibnamefont
  {Astrakharchik}}, \bibinfo {author} {\bibfnamefont {B.~A.}\ \bibnamefont
  {Malomed}}, \ and\ \bibinfo {author} {\bibfnamefont {D.~S.}\ \bibnamefont
  {Petrov}},\ }\bibfield  {title} {\bibinfo {title} {\emph {Collective
  excitations of a one-dimensional quantum droplet}},\ }\href {\doibase
  10.1103/PhysRevA.101.051601} {\bibfield  {journal} {\bibinfo  {journal}
  {Phys. Rev. A}\ }\textbf {\bibinfo {volume} {101}},\ \bibinfo {pages}
  {051601(R)} (\bibinfo {year} {2020})}\BibitemShut {NoStop}%
\bibitem [{\citenamefont {Andrews}\ \emph {et~al.}(1997)\citenamefont
  {Andrews}, \citenamefont {Kurn}, \citenamefont {Miesner}, \citenamefont
  {Durfee}, \citenamefont {Townsend}, \citenamefont {Inouye},\ and\
  \citenamefont {Ketterle}}]{Andrews1997}%
  \BibitemOpen
  \bibfield  {author} {\bibinfo {author} {\bibfnamefont {M.~R.}\ \bibnamefont
  {Andrews}}, \bibinfo {author} {\bibfnamefont {D.~M.}\ \bibnamefont {Kurn}},
  \bibinfo {author} {\bibfnamefont {H.-J.}\ \bibnamefont {Miesner}}, \bibinfo
  {author} {\bibfnamefont {D.~S.}\ \bibnamefont {Durfee}}, \bibinfo {author}
  {\bibfnamefont {C.~G.}\ \bibnamefont {Townsend}}, \bibinfo {author}
  {\bibfnamefont {S.}~\bibnamefont {Inouye}}, \ and\ \bibinfo {author}
  {\bibfnamefont {W.}~\bibnamefont {Ketterle}},\ }\bibfield  {title} {\bibinfo
  {title} {\emph {Propagation of Sound in a Bose-Einstein Condensate}},\ }\href
  {\doibase 10.1103/PhysRevLett.79.553} {\bibfield  {journal} {\bibinfo
  {journal} {Phys. Rev. Lett.}\ }\textbf {\bibinfo {volume} {79}},\ \bibinfo
  {pages} {553} (\bibinfo {year} {1997})}\BibitemShut {NoStop}%
\bibitem [{\citenamefont {Joseph}\ \emph {et~al.}(2007)\citenamefont {Joseph},
  \citenamefont {Clancy}, \citenamefont {Luo}, \citenamefont {Kinast},
  \citenamefont {Turlapov},\ and\ \citenamefont {Thomas}}]{Joseph2007}%
  \BibitemOpen
  \bibfield  {author} {\bibinfo {author} {\bibfnamefont {J.}~\bibnamefont
  {Joseph}}, \bibinfo {author} {\bibfnamefont {B.}~\bibnamefont {Clancy}},
  \bibinfo {author} {\bibfnamefont {L.}~\bibnamefont {Luo}}, \bibinfo {author}
  {\bibfnamefont {J.}~\bibnamefont {Kinast}}, \bibinfo {author} {\bibfnamefont
  {A.}~\bibnamefont {Turlapov}}, \ and\ \bibinfo {author} {\bibfnamefont
  {J.~E.}\ \bibnamefont {Thomas}},\ }\bibfield  {title} {\bibinfo {title}
  {\emph {Measurement of Sound Velocity in a Fermi Gas near a Feshbach
  Resonance}},\ }\href {\doibase 10.1103/PhysRevLett.98.170401} {\bibfield
  {journal} {\bibinfo  {journal} {Phys. Rev. Lett.}\ }\textbf {\bibinfo
  {volume} {98}},\ \bibinfo {pages} {170401} (\bibinfo {year}
  {2007})}\BibitemShut {NoStop}%
\bibitem [{\citenamefont {Ho}\ and\ \citenamefont {Zhou}(2009)}]{Ho2009}%
  \BibitemOpen
  \bibfield  {author} {\bibinfo {author} {\bibfnamefont {T.-L.}\ \bibnamefont
  {Ho}}\ and\ \bibinfo {author} {\bibfnamefont {Q.}~\bibnamefont {Zhou}},\
  }\bibfield  {title} {\bibinfo {title} {\emph {Obtaining the phase diagram and
  thermodynamic quantities of bulk systems from the densities of trapped
  gases}},\ }\href {https://doi.org/10.1038/nphys1477} {\bibfield  {journal}
  {\bibinfo  {journal} {Nature Phys.}\ }\textbf {\bibinfo {volume} {6}},\
  \bibinfo {pages} {131 EP } (\bibinfo {year} {2009})}\BibitemShut {NoStop}%
\bibitem [{\citenamefont {Nascimb{\`{e}}ne}\ \emph {et~al.}(2010)\citenamefont
  {Nascimb{\`{e}}ne}, \citenamefont {Navon}, \citenamefont {Jiang},
  \citenamefont {Chevy},\ and\ \citenamefont {Salomon}}]{Nascimbene2010b}%
  \BibitemOpen
  \bibfield  {author} {\bibinfo {author} {\bibfnamefont {S.}~\bibnamefont
  {Nascimb{\`{e}}ne}}, \bibinfo {author} {\bibfnamefont {N.}~\bibnamefont
  {Navon}}, \bibinfo {author} {\bibfnamefont {K.~J.}\ \bibnamefont {Jiang}},
  \bibinfo {author} {\bibfnamefont {F.}~\bibnamefont {Chevy}}, \ and\ \bibinfo
  {author} {\bibfnamefont {C.}~\bibnamefont {Salomon}},\ }\bibfield  {title}
  {\bibinfo {title} {\emph {{Exploring the thermodynamics of a universal Fermi
  gas}}},\ }\href {https://www.nature.com/articles/nature08814} {\bibfield
  {journal} {\bibinfo  {journal} {Nature}\ }\textbf {\bibinfo {volume} {463}},\
  \bibinfo {pages} {1057} (\bibinfo {year} {2010})}\BibitemShut {NoStop}%
\bibitem [{\citenamefont {Ku}\ \emph {et~al.}(2012)\citenamefont {Ku},
  \citenamefont {Sommer}, \citenamefont {Cheuk},\ and\ \citenamefont
  {Zwierlein}}]{Ku2012}%
  \BibitemOpen
  \bibfield  {author} {\bibinfo {author} {\bibfnamefont {M.~J.~H.}\
  \bibnamefont {Ku}}, \bibinfo {author} {\bibfnamefont {A.~T.}\ \bibnamefont
  {Sommer}}, \bibinfo {author} {\bibfnamefont {L.~W.}\ \bibnamefont {Cheuk}}, \
  and\ \bibinfo {author} {\bibfnamefont {M.~W.}\ \bibnamefont {Zwierlein}},\
  }\bibfield  {title} {\bibinfo {title} {\emph {Revealing the Superfluid Lambda
  Transition in the Universal Thermodynamics of a Unitary Fermi Gas}},\ }\href
  {\doibase 10.1126/science.1214987} {\bibfield  {journal} {\bibinfo  {journal}
  {Science}\ }\textbf {\bibinfo {volume} {335}},\ \bibinfo {pages} {563}
  (\bibinfo {year} {2012})}\BibitemShut {NoStop}%
\bibitem [{\citenamefont {Hoinka}\ \emph {et~al.}(2013)\citenamefont {Hoinka},
  \citenamefont {Lingham}, \citenamefont {Fenech}, \citenamefont {Hu},
  \citenamefont {Vale}, \citenamefont {Drut},\ and\ \citenamefont
  {Gandolfi}}]{Hoinka2013}%
  \BibitemOpen
  \bibfield  {author} {\bibinfo {author} {\bibfnamefont {S.}~\bibnamefont
  {Hoinka}}, \bibinfo {author} {\bibfnamefont {M.}~\bibnamefont {Lingham}},
  \bibinfo {author} {\bibfnamefont {K.}~\bibnamefont {Fenech}}, \bibinfo
  {author} {\bibfnamefont {H.}~\bibnamefont {Hu}}, \bibinfo {author}
  {\bibfnamefont {C.~J.}\ \bibnamefont {Vale}}, \bibinfo {author}
  {\bibfnamefont {J.~E.}\ \bibnamefont {Drut}}, \ and\ \bibinfo {author}
  {\bibfnamefont {S.}~\bibnamefont {Gandolfi}},\ }\bibfield  {title} {\bibinfo
  {title} {\emph {Precise Determination of the Structure Factor and Contact in
  a Unitary Fermi Gas}},\ }\href {\doibase 10.1103/PhysRevLett.110.055305}
  {\bibfield  {journal} {\bibinfo  {journal} {Phys. Rev. Lett.}\ }\textbf
  {\bibinfo {volume} {110}},\ \bibinfo {pages} {055305} (\bibinfo {year}
  {2013})}\BibitemShut {NoStop}%
\bibitem [{\citenamefont {Salces-Carcoba}\ \emph {et~al.}(2018)\citenamefont
  {Salces-Carcoba}, \citenamefont {Billington}, \citenamefont {Putra},
  \citenamefont {Yue}, \citenamefont {Sugawa},\ and\ \citenamefont
  {Spielman}}]{Salces-Carcoba2018}%
  \BibitemOpen
  \bibfield  {author} {\bibinfo {author} {\bibfnamefont {F.}~\bibnamefont
  {Salces-Carcoba}}, \bibinfo {author} {\bibfnamefont {C.~J.}\ \bibnamefont
  {Billington}}, \bibinfo {author} {\bibfnamefont {A.}~\bibnamefont {Putra}},
  \bibinfo {author} {\bibfnamefont {Y.}~\bibnamefont {Yue}}, \bibinfo {author}
  {\bibfnamefont {S.}~\bibnamefont {Sugawa}}, \ and\ \bibinfo {author}
  {\bibfnamefont {I.~B.}\ \bibnamefont {Spielman}},\ }\bibfield  {title}
  {\bibinfo {title} {\emph {Equations of state from individual one-dimensional
  Bose gases}},\ }\href {http://stacks.iop.org/1367-2630/20/i=11/a=113032}
  {\bibfield  {journal} {\bibinfo  {journal} {New J. Phys.}\ }\textbf {\bibinfo
  {volume} {20}},\ \bibinfo {pages} {113032} (\bibinfo {year}
  {2018})}\BibitemShut {NoStop}%
\bibitem [{\citenamefont {Wang}\ \emph {et~al.}(2020)\citenamefont {Wang},
  \citenamefont {Hu},\ and\ \citenamefont {Liu}}]{Wang2020}%
  \BibitemOpen
  \bibfield  {author} {\bibinfo {author} {\bibfnamefont {J.}~\bibnamefont
  {Wang}}, \bibinfo {author} {\bibfnamefont {H.}~\bibnamefont {Hu}}, \ and\
  \bibinfo {author} {\bibfnamefont {X.-J.}\ \bibnamefont {Liu}},\ }\bibfield
  {title} {\bibinfo {title} {\emph {Thermal destabilization of self-bound
  ultradilute quantum droplets}},\ }\href
  {https://doi.org/10.1088/1367-2630/abbe55} {\bibfield  {journal} {\bibinfo
  {journal} {New J. Phys.}\ }\textbf {\bibinfo {volume} {22}},\ \bibinfo
  {pages} {103044} (\bibinfo {year} {2020})}\BibitemShut {NoStop}%
\bibitem [{\citenamefont {Ota}\ and\ \citenamefont
  {Astrakharchik}(2020)}]{Ota2020}%
  \BibitemOpen
  \bibfield  {author} {\bibinfo {author} {\bibfnamefont {M.}~\bibnamefont
  {Ota}}\ and\ \bibinfo {author} {\bibfnamefont {G.~E.}\ \bibnamefont
  {Astrakharchik}},\ }\bibfield  {title} {\bibinfo {title} {\emph {{Beyond
  Lee-Huang-Yang description of self-bound Bose mixtures}}},\ }\href {\doibase
  10.21468/SciPostPhys.9.2.020} {\bibfield  {journal} {\bibinfo  {journal}
  {SciPost Phys.}\ }\textbf {\bibinfo {volume} {9}},\ \bibinfo {pages} {20}
  (\bibinfo {year} {2020})}\BibitemShut {NoStop}%
\bibitem [{\citenamefont {Meinert}\ \emph {et~al.}(2015)\citenamefont
  {Meinert}, \citenamefont {Panfil}, \citenamefont {Mark}, \citenamefont
  {Lauber}, \citenamefont {Caux},\ and\ \citenamefont
  {N\"agerl}}]{Meinert2015}%
  \BibitemOpen
  \bibfield  {author} {\bibinfo {author} {\bibfnamefont {F.}~\bibnamefont
  {Meinert}}, \bibinfo {author} {\bibfnamefont {M.}~\bibnamefont {Panfil}},
  \bibinfo {author} {\bibfnamefont {M.~J.}\ \bibnamefont {Mark}}, \bibinfo
  {author} {\bibfnamefont {K.}~\bibnamefont {Lauber}}, \bibinfo {author}
  {\bibfnamefont {J.-S.}\ \bibnamefont {Caux}}, \ and\ \bibinfo {author}
  {\bibfnamefont {H.-C.}\ \bibnamefont {N\"agerl}},\ }\bibfield  {title}
  {\bibinfo {title} {\emph {Probing the Excitations of a Lieb-Liniger Gas from
  Weak to Strong Coupling}},\ }\href {\doibase 10.1103/PhysRevLett.115.085301}
  {\bibfield  {journal} {\bibinfo  {journal} {Phys. Rev. Lett.}\ }\textbf
  {\bibinfo {volume} {115}},\ \bibinfo {pages} {085301} (\bibinfo {year}
  {2015})}\BibitemShut {NoStop}%
\bibitem [{\citenamefont {Fabbri}\ \emph {et~al.}(2015)\citenamefont {Fabbri},
  \citenamefont {Panfil}, \citenamefont {Cl\'ement}, \citenamefont {Fallani},
  \citenamefont {Inguscio}, \citenamefont {Fort},\ and\ \citenamefont
  {Caux}}]{Fabbri2015}%
  \BibitemOpen
  \bibfield  {author} {\bibinfo {author} {\bibfnamefont {N.}~\bibnamefont
  {Fabbri}}, \bibinfo {author} {\bibfnamefont {M.}~\bibnamefont {Panfil}},
  \bibinfo {author} {\bibfnamefont {D.}~\bibnamefont {Cl\'ement}}, \bibinfo
  {author} {\bibfnamefont {L.}~\bibnamefont {Fallani}}, \bibinfo {author}
  {\bibfnamefont {M.}~\bibnamefont {Inguscio}}, \bibinfo {author}
  {\bibfnamefont {C.}~\bibnamefont {Fort}}, \ and\ \bibinfo {author}
  {\bibfnamefont {J.-S.}\ \bibnamefont {Caux}},\ }\bibfield  {title} {\bibinfo
  {title} {\emph {Dynamical structure factor of one-dimensional Bose gases:
  Experimental signatures of beyond-Luttinger-liquid physics}},\ }\href
  {\doibase 10.1103/PhysRevA.91.043617} {\bibfield  {journal} {\bibinfo
  {journal} {Phys. Rev. A}\ }\textbf {\bibinfo {volume} {91}},\ \bibinfo
  {pages} {043617} (\bibinfo {year} {2015})}\BibitemShut {NoStop}%
\bibitem [{\citenamefont {De~Rosi}\ \emph {et~al.}(2019)\citenamefont
  {De~Rosi}, \citenamefont {Massignan}, \citenamefont {Lewenstein},\ and\
  \citenamefont {Astrakharchik}}]{DeRosi2019}%
  \BibitemOpen
  \bibfield  {author} {\bibinfo {author} {\bibfnamefont {G.}~\bibnamefont
  {De~Rosi}}, \bibinfo {author} {\bibfnamefont {P.}~\bibnamefont {Massignan}},
  \bibinfo {author} {\bibfnamefont {M.}~\bibnamefont {Lewenstein}}, \ and\
  \bibinfo {author} {\bibfnamefont {G.~E.}\ \bibnamefont {Astrakharchik}},\
  }\bibfield  {title} {\bibinfo {title} {\emph {Beyond-Luttinger-liquid
  thermodynamics of a one-dimensional Bose gas with repulsive contact
  interactions}},\ }\href {\doibase 10.1103/PhysRevResearch.1.033083}
  {\bibfield  {journal} {\bibinfo  {journal} {Phys. Rev. Res.}\ }\textbf
  {\bibinfo {volume} {1}},\ \bibinfo {pages} {033083} (\bibinfo {year}
  {2019})}\BibitemShut {NoStop}%
\bibitem [{\citenamefont {De~Rosi}\ \emph {et~al.}(2017)\citenamefont
  {De~Rosi}, \citenamefont {Astrakharchik},\ and\ \citenamefont
  {Stringari}}]{DeRosi2017}%
  \BibitemOpen
  \bibfield  {author} {\bibinfo {author} {\bibfnamefont {G.}~\bibnamefont
  {De~Rosi}}, \bibinfo {author} {\bibfnamefont {G.~E.}\ \bibnamefont
  {Astrakharchik}}, \ and\ \bibinfo {author} {\bibfnamefont {S.}~\bibnamefont
  {Stringari}},\ }\bibfield  {title} {\bibinfo {title} {\emph {Thermodynamic
  behavior of a one-dimensional Bose gas at low temperature}},\ }\href
  {\doibase 10.1103/PhysRevA.96.013613} {\bibfield  {journal} {\bibinfo
  {journal} {Phys. Rev. A}\ }\textbf {\bibinfo {volume} {96}},\ \bibinfo
  {pages} {013613} (\bibinfo {year} {2017})}\BibitemShut {NoStop}%
\bibitem [{\citenamefont {Errico}\ \emph {et~al.}(2007)\citenamefont {Errico},
  \citenamefont {Zaccanti}, \citenamefont {Fattori}, \citenamefont {Roati},
  \citenamefont {Inguscio}, \citenamefont {Modugno},\ and\ \citenamefont
  {Simoni}}]{DErrico2007}%
  \BibitemOpen
  \bibfield  {author} {\bibinfo {author} {\bibfnamefont {C.~D.}\ \bibnamefont
  {Errico}}, \bibinfo {author} {\bibfnamefont {M.}~\bibnamefont {Zaccanti}},
  \bibinfo {author} {\bibfnamefont {M.}~\bibnamefont {Fattori}}, \bibinfo
  {author} {\bibfnamefont {G.}~\bibnamefont {Roati}}, \bibinfo {author}
  {\bibfnamefont {M.}~\bibnamefont {Inguscio}}, \bibinfo {author}
  {\bibfnamefont {G.}~\bibnamefont {Modugno}}, \ and\ \bibinfo {author}
  {\bibfnamefont {A.}~\bibnamefont {Simoni}},\ }\bibfield  {title} {\bibinfo
  {title} {\emph {Feshbach resonances in ultracold 39K}},\ }\href {\doibase
  10.1088/1367-2630/9/7/223} {\bibfield  {journal} {\bibinfo  {journal} {New J.
  Phys.}\ }\textbf {\bibinfo {volume} {9}},\ \bibinfo {pages} {223} (\bibinfo
  {year} {2007})}\BibitemShut {NoStop}%
\bibitem [{\citenamefont {Chin}\ \emph {et~al.}(2010)\citenamefont {Chin},
  \citenamefont {Grimm}, \citenamefont {Julienne},\ and\ \citenamefont
  {Tiesinga}}]{Chin2010}%
  \BibitemOpen
  \bibfield  {author} {\bibinfo {author} {\bibfnamefont {C.}~\bibnamefont
  {Chin}}, \bibinfo {author} {\bibfnamefont {R.}~\bibnamefont {Grimm}},
  \bibinfo {author} {\bibfnamefont {P.}~\bibnamefont {Julienne}}, \ and\
  \bibinfo {author} {\bibfnamefont {E.}~\bibnamefont {Tiesinga}},\ }\bibfield
  {title} {\bibinfo {title} {\emph {Feshbach resonances in ultracold gases}},\
  }\href {\doibase 10.1103/RevModPhys.82.1225} {\bibfield  {journal} {\bibinfo
  {journal} {Rev. Mod. Phys.}\ }\textbf {\bibinfo {volume} {82}},\ \bibinfo
  {pages} {1225} (\bibinfo {year} {2010})}\BibitemShut {NoStop}%
\bibitem [{\citenamefont {Tanzi}\ \emph {et~al.}(2018)\citenamefont {Tanzi},
  \citenamefont {Cabrera}, \citenamefont {Sanz}, \citenamefont {Cheiney},
  \citenamefont {Tomza},\ and\ \citenamefont {Tarruell}}]{Tanzi2018}%
  \BibitemOpen
  \bibfield  {author} {\bibinfo {author} {\bibfnamefont {L.}~\bibnamefont
  {Tanzi}}, \bibinfo {author} {\bibfnamefont {C.~R.}\ \bibnamefont {Cabrera}},
  \bibinfo {author} {\bibfnamefont {J.}~\bibnamefont {Sanz}}, \bibinfo {author}
  {\bibfnamefont {P.}~\bibnamefont {Cheiney}}, \bibinfo {author} {\bibfnamefont
  {M.}~\bibnamefont {Tomza}}, \ and\ \bibinfo {author} {\bibfnamefont
  {L.}~\bibnamefont {Tarruell}},\ }\bibfield  {title} {\bibinfo {title} {\emph
  {Feshbach resonances in potassium Bose-Bose mixtures}},\ }\href {\doibase
  10.1103/PhysRevA.98.062712} {\bibfield  {journal} {\bibinfo  {journal} {Phys.
  Rev. A}\ }\textbf {\bibinfo {volume} {98}},\ \bibinfo {pages} {062712}
  (\bibinfo {year} {2018})}\BibitemShut {NoStop}%
\bibitem [{\citenamefont {Lieb}\ and\ \citenamefont
  {Liniger}(1963)}]{Lieb1963}%
  \BibitemOpen
  \bibfield  {author} {\bibinfo {author} {\bibfnamefont {E.~H.}\ \bibnamefont
  {Lieb}}\ and\ \bibinfo {author} {\bibfnamefont {W.}~\bibnamefont {Liniger}},\
  }\bibfield  {title} {\bibinfo {title} {\emph {Exact Analysis of an
  Interacting Bose Gas. I. The General Solution and the Ground State}},\ }\href
  {\doibase 10.1103/PhysRev.130.1605} {\bibfield  {journal} {\bibinfo
  {journal} {Phys. Rev.}\ }\textbf {\bibinfo {volume} {130}},\ \bibinfo {pages}
  {1605} (\bibinfo {year} {1963})}\BibitemShut {NoStop}%
\bibitem [{\citenamefont {De~Rosi}\ and\ \citenamefont
  {Stringari}(2015)}]{DeRosi2015}%
  \BibitemOpen
  \bibfield  {author} {\bibinfo {author} {\bibfnamefont {G.}~\bibnamefont
  {De~Rosi}}\ and\ \bibinfo {author} {\bibfnamefont {S.}~\bibnamefont
  {Stringari}},\ }\bibfield  {title} {\bibinfo {title} {\emph {Collective
  oscillations of a trapped quantum gas in low dimensions}},\ }\href {\doibase
  10.1103/PhysRevA.92.053617} {\bibfield  {journal} {\bibinfo  {journal} {Phys.
  Rev. A}\ }\textbf {\bibinfo {volume} {92}},\ \bibinfo {pages} {053617}
  (\bibinfo {year} {2015})}\BibitemShut {NoStop}%
\bibitem [{\citenamefont {De~Rosi}\ and\ \citenamefont
  {Stringari}(2016)}]{DeRosi2016}%
  \BibitemOpen
  \bibfield  {author} {\bibinfo {author} {\bibfnamefont {G.}~\bibnamefont
  {De~Rosi}}\ and\ \bibinfo {author} {\bibfnamefont {S.}~\bibnamefont
  {Stringari}},\ }\bibfield  {title} {\bibinfo {title} {\emph {Hydrodynamic
  versus collisionless dynamics of a one-dimensional harmonically trapped Bose
  gas}},\ }\href {\doibase 10.1103/PhysRevA.94.063605} {\bibfield  {journal}
  {\bibinfo  {journal} {Phys. Rev. A}\ }\textbf {\bibinfo {volume} {94}},\
  \bibinfo {pages} {063605} (\bibinfo {year} {2016})}\BibitemShut {NoStop}%
\bibitem [{\citenamefont {Barth}\ and\ \citenamefont
  {Zwerger}(2011)}]{Barth2011}%
  \BibitemOpen
  \bibfield  {author} {\bibinfo {author} {\bibfnamefont {M.}~\bibnamefont
  {Barth}}\ and\ \bibinfo {author} {\bibfnamefont {W.}~\bibnamefont
  {Zwerger}},\ }\bibfield  {title} {\bibinfo {title} {\emph {Tan relations in
  one dimension}},\ }\href {\doibase https://doi.org/10.1016/j.aop.2011.05.010}
  {\bibfield  {journal} {\bibinfo  {journal} {Ann. Phys.}\ }\textbf {\bibinfo
  {volume} {326}},\ \bibinfo {pages} {2544 } (\bibinfo {year}
  {2011})}\BibitemShut {NoStop}%
\bibitem [{\citenamefont {Wild}\ \emph {et~al.}(2012)\citenamefont {Wild},
  \citenamefont {Makotyn}, \citenamefont {Pino}, \citenamefont {Cornell},\ and\
  \citenamefont {Jin}}]{Wild2012}%
  \BibitemOpen
  \bibfield  {author} {\bibinfo {author} {\bibfnamefont {R.~J.}\ \bibnamefont
  {Wild}}, \bibinfo {author} {\bibfnamefont {P.}~\bibnamefont {Makotyn}},
  \bibinfo {author} {\bibfnamefont {J.~M.}\ \bibnamefont {Pino}}, \bibinfo
  {author} {\bibfnamefont {E.~A.}\ \bibnamefont {Cornell}}, \ and\ \bibinfo
  {author} {\bibfnamefont {D.~S.}\ \bibnamefont {Jin}},\ }\bibfield  {title}
  {\bibinfo {title} {\emph {Measurements of Tan's Contact in an Atomic
  Bose-Einstein Condensate}},\ }\href {\doibase 10.1103/PhysRevLett.108.145305}
  {\bibfield  {journal} {\bibinfo  {journal} {Phys. Rev. Lett.}\ }\textbf
  {\bibinfo {volume} {108}},\ \bibinfo {pages} {145305} (\bibinfo {year}
  {2012})}\BibitemShut {NoStop}%
\bibitem [{\citenamefont {Olshanii}\ and\ \citenamefont
  {Dunjko}(2003)}]{Olshanii2003}%
  \BibitemOpen
  \bibfield  {author} {\bibinfo {author} {\bibfnamefont {M.}~\bibnamefont
  {Olshanii}}\ and\ \bibinfo {author} {\bibfnamefont {V.}~\bibnamefont
  {Dunjko}},\ }\bibfield  {title} {\bibinfo {title} {\emph {Short-Distance
  Correlation Properties of the Lieb-Liniger System and Momentum Distributions
  of Trapped One-Dimensional Atomic Gases}},\ }\href {\doibase
  10.1103/PhysRevLett.91.090401} {\bibfield  {journal} {\bibinfo  {journal}
  {Phys. Rev. Lett.}\ }\textbf {\bibinfo {volume} {91}},\ \bibinfo {pages}
  {090401} (\bibinfo {year} {2003})}\BibitemShut {NoStop}%
\bibitem [{\citenamefont {Tan}(2008{\natexlab{a}})}]{Tan2008}%
  \BibitemOpen
  \bibfield  {author} {\bibinfo {author} {\bibfnamefont {S.}~\bibnamefont
  {Tan}},\ }\bibfield  {title} {\bibinfo {title} {\emph {Energetics of a
  strongly correlated Fermi gas}},\ }\href {\doibase
  https://doi.org/10.1016/j.aop.2008.03.004} {\bibfield  {journal} {\bibinfo
  {journal} {Ann. Phys.}\ }\textbf {\bibinfo {volume} {323}},\ \bibinfo {pages}
  {2952 } (\bibinfo {year} {2008}{\natexlab{a}})}\BibitemShut {NoStop}%
\bibitem [{\citenamefont {Tan}(2008{\natexlab{b}})}]{Tan22008}%
  \BibitemOpen
  \bibfield  {author} {\bibinfo {author} {\bibfnamefont {S.}~\bibnamefont
  {Tan}},\ }\bibfield  {title} {\bibinfo {title} {\emph {Generalized virial
  theorem and pressure relation for a strongly correlated Fermi gas}},\ }\href
  {\doibase https://doi.org/10.1016/j.aop.2008.03.003} {\bibfield  {journal}
  {\bibinfo  {journal} {Ann. Phys.}\ }\textbf {\bibinfo {volume} {323}},\
  \bibinfo {pages} {2987 } (\bibinfo {year} {2008}{\natexlab{b}})}\BibitemShut
  {NoStop}%
\bibitem [{\citenamefont {Tan}(2008{\natexlab{c}})}]{Tan32008}%
  \BibitemOpen
  \bibfield  {author} {\bibinfo {author} {\bibfnamefont {S.}~\bibnamefont
  {Tan}},\ }\bibfield  {title} {\bibinfo {title} {\emph {Large momentum part of
  a strongly correlated Fermi gas}},\ }\href {\doibase
  https://doi.org/10.1016/j.aop.2008.03.005} {\bibfield  {journal} {\bibinfo
  {journal} {Ann. Phys.}\ }\textbf {\bibinfo {volume} {323}},\ \bibinfo {pages}
  {2971 } (\bibinfo {year} {2008}{\natexlab{c}})}\BibitemShut {NoStop}%
\bibitem [{\citenamefont {Braaten}\ \emph {et~al.}(2011)\citenamefont
  {Braaten}, \citenamefont {Kang},\ and\ \citenamefont
  {Platter}}]{Braaten2011}%
  \BibitemOpen
  \bibfield  {author} {\bibinfo {author} {\bibfnamefont {E.}~\bibnamefont
  {Braaten}}, \bibinfo {author} {\bibfnamefont {D.}~\bibnamefont {Kang}}, \
  and\ \bibinfo {author} {\bibfnamefont {L.}~\bibnamefont {Platter}},\
  }\bibfield  {title} {\bibinfo {title} {\emph {Universal Relations for
  Identical Bosons from Three-Body Physics}},\ }\href {\doibase
  10.1103/PhysRevLett.106.153005} {\bibfield  {journal} {\bibinfo  {journal}
  {Phys. Rev. Lett.}\ }\textbf {\bibinfo {volume} {106}},\ \bibinfo {pages}
  {153005} (\bibinfo {year} {2011})}\BibitemShut {NoStop}%
\bibitem [{\citenamefont {Yao}\ \emph {et~al.}(2018)\citenamefont {Yao},
  \citenamefont {Cl\'ement}, \citenamefont {Minguzzi}, \citenamefont
  {Vignolo},\ and\ \citenamefont {Sanchez-Palencia}}]{Yao2018}%
  \BibitemOpen
  \bibfield  {author} {\bibinfo {author} {\bibfnamefont {H.}~\bibnamefont
  {Yao}}, \bibinfo {author} {\bibfnamefont {D.}~\bibnamefont {Cl\'ement}},
  \bibinfo {author} {\bibfnamefont {A.}~\bibnamefont {Minguzzi}}, \bibinfo
  {author} {\bibfnamefont {P.}~\bibnamefont {Vignolo}}, \ and\ \bibinfo
  {author} {\bibfnamefont {L.}~\bibnamefont {Sanchez-Palencia}},\ }\bibfield
  {title} {\bibinfo {title} {\emph {Tan's Contact for Trapped Lieb-Liniger
  Bosons at Finite Temperature}},\ }\href {\doibase
  10.1103/PhysRevLett.121.220402} {\bibfield  {journal} {\bibinfo  {journal}
  {Phys. Rev. Lett.}\ }\textbf {\bibinfo {volume} {121}},\ \bibinfo {pages}
  {220402} (\bibinfo {year} {2018})}\BibitemShut {NoStop}%
\bibitem [{\citenamefont {Fetter}\ and\ \citenamefont
  {Walecka}(1971)}]{FetterBook}%
  \BibitemOpen
  \bibfield  {author} {\bibinfo {author} {\bibfnamefont {A.}~\bibnamefont
  {Fetter}}\ and\ \bibinfo {author} {\bibfnamefont {J.}~\bibnamefont
  {Walecka}},\ }\href@noop {} {\emph {\bibinfo {title} {Quantum Theory of
  Many-Particle Systems}}},\ Dover Books on Physics Series\ (\bibinfo
  {publisher} {Dover, New York},\ \bibinfo {year} {1971})\BibitemShut {NoStop}%
\bibitem [{\citenamefont {P\^a\ifmmode~\mbox{\c{t}}\else \c{t}\fi{}u}\ and\
  \citenamefont {Kl\"umper}(2017)}]{Patu2017}%
  \BibitemOpen
  \bibfield  {author} {\bibinfo {author} {\bibfnamefont {O.~I.}\ \bibnamefont
  {P\^a\ifmmode~\mbox{\c{t}}\else \c{t}\fi{}u}}\ and\ \bibinfo {author}
  {\bibfnamefont {A.}~\bibnamefont {Kl\"umper}},\ }\bibfield  {title} {\bibinfo
  {title} {\emph {Universal Tan relations for quantum gases in one
  dimension}},\ }\href {\doibase 10.1103/PhysRevA.96.063612} {\bibfield
  {journal} {\bibinfo  {journal} {Phys. Rev. A}\ }\textbf {\bibinfo {volume}
  {96}},\ \bibinfo {pages} {063612} (\bibinfo {year} {2017})}\BibitemShut
  {NoStop}%
\bibitem [{\citenamefont {Pitaevskii}\ and\ \citenamefont
  {Rosch}(1997)}]{Pitaevskii1997}%
  \BibitemOpen
  \bibfield  {author} {\bibinfo {author} {\bibfnamefont {L.~P.}\ \bibnamefont
  {Pitaevskii}}\ and\ \bibinfo {author} {\bibfnamefont {A.}~\bibnamefont
  {Rosch}},\ }\bibfield  {title} {\bibinfo {title} {\emph {Breathing modes and
  hidden symmetry of trapped atoms in two dimensions}},\ }\href {\doibase
  10.1103/PhysRevA.55.R853} {\bibfield  {journal} {\bibinfo  {journal} {Phys.
  Rev. A}\ }\textbf {\bibinfo {volume} {55}},\ \bibinfo {pages} {R853}
  (\bibinfo {year} {1997})}\BibitemShut {NoStop}%
\bibitem [{\citenamefont {Olshanii}\ \emph {et~al.}(2010)\citenamefont
  {Olshanii}, \citenamefont {Perrin},\ and\ \citenamefont
  {Lorent}}]{Olshanii2010}%
  \BibitemOpen
  \bibfield  {author} {\bibinfo {author} {\bibfnamefont {M.}~\bibnamefont
  {Olshanii}}, \bibinfo {author} {\bibfnamefont {H.}~\bibnamefont {Perrin}}, \
  and\ \bibinfo {author} {\bibfnamefont {V.}~\bibnamefont {Lorent}},\
  }\bibfield  {title} {\bibinfo {title} {\emph {Example of a Quantum Anomaly in
  the Physics of Ultracold Gases}},\ }\href {\doibase
  10.1103/PhysRevLett.105.095302} {\bibfield  {journal} {\bibinfo  {journal}
  {Phys. Rev. Lett.}\ }\textbf {\bibinfo {volume} {105}},\ \bibinfo {pages}
  {095302} (\bibinfo {year} {2010})}\BibitemShut {NoStop}%
\bibitem [{\citenamefont {Holten}\ \emph {et~al.}(2018)\citenamefont {Holten},
  \citenamefont {Bayha}, \citenamefont {Klein}, \citenamefont {Murthy},
  \citenamefont {Preiss},\ and\ \citenamefont {Jochim}}]{Holten2018}%
  \BibitemOpen
  \bibfield  {author} {\bibinfo {author} {\bibfnamefont {M.}~\bibnamefont
  {Holten}}, \bibinfo {author} {\bibfnamefont {L.}~\bibnamefont {Bayha}},
  \bibinfo {author} {\bibfnamefont {A.~C.}\ \bibnamefont {Klein}}, \bibinfo
  {author} {\bibfnamefont {P.~A.}\ \bibnamefont {Murthy}}, \bibinfo {author}
  {\bibfnamefont {P.~M.}\ \bibnamefont {Preiss}}, \ and\ \bibinfo {author}
  {\bibfnamefont {S.}~\bibnamefont {Jochim}},\ }\bibfield  {title} {\bibinfo
  {title} {\emph {Anomalous Breaking of Scale Invariance in a Two-Dimensional
  Fermi Gas}},\ }\href {\doibase 10.1103/PhysRevLett.121.120401} {\bibfield
  {journal} {\bibinfo  {journal} {Phys. Rev. Lett.}\ }\textbf {\bibinfo
  {volume} {121}},\ \bibinfo {pages} {120401} (\bibinfo {year}
  {2018})}\BibitemShut {NoStop}%
\bibitem [{\citenamefont {Peppler}\ \emph {et~al.}(2018)\citenamefont
  {Peppler}, \citenamefont {Dyke}, \citenamefont {Zamorano}, \citenamefont
  {Herrera}, \citenamefont {Hoinka},\ and\ \citenamefont {Vale}}]{Peppler2018}%
  \BibitemOpen
  \bibfield  {author} {\bibinfo {author} {\bibfnamefont {T.}~\bibnamefont
  {Peppler}}, \bibinfo {author} {\bibfnamefont {P.}~\bibnamefont {Dyke}},
  \bibinfo {author} {\bibfnamefont {M.}~\bibnamefont {Zamorano}}, \bibinfo
  {author} {\bibfnamefont {I.}~\bibnamefont {Herrera}}, \bibinfo {author}
  {\bibfnamefont {S.}~\bibnamefont {Hoinka}}, \ and\ \bibinfo {author}
  {\bibfnamefont {C.~J.}\ \bibnamefont {Vale}},\ }\bibfield  {title} {\bibinfo
  {title} {\emph {Quantum Anomaly and 2D-3D Crossover in Strongly Interacting
  Fermi Gases}},\ }\href {\doibase 10.1103/PhysRevLett.121.120402} {\bibfield
  {journal} {\bibinfo  {journal} {Phys. Rev. Lett.}\ }\textbf {\bibinfo
  {volume} {121}},\ \bibinfo {pages} {120402} (\bibinfo {year}
  {2018})}\BibitemShut {NoStop}%
\bibitem [{\citenamefont {Pethick}\ and\ \citenamefont
  {Smith}(2008)}]{Pethick2008}%
  \BibitemOpen
  \bibfield  {author} {\bibinfo {author} {\bibfnamefont {C.~J.}\ \bibnamefont
  {Pethick}}\ and\ \bibinfo {author} {\bibfnamefont {H.}~\bibnamefont
  {Smith}},\ }\href@noop {} {\emph {\bibinfo {title} {Bose-Einstein
  Condensation in Dilute Gases}}},\ \bibinfo {edition} {2nd}\ ed.\ (\bibinfo
  {publisher} {Cambridge University Press, Cambridge, UK},\ \bibinfo {year}
  {2008})\BibitemShut {NoStop}%
\bibitem [{\citenamefont {Lieb}(1963)}]{Lieb21963}%
  \BibitemOpen
  \bibfield  {author} {\bibinfo {author} {\bibfnamefont {E.~H.}\ \bibnamefont
  {Lieb}},\ }\bibfield  {title} {\bibinfo {title} {\emph {Exact Analysis of an
  Interacting Bose Gas. II. The Excitation Spectrum}},\ }\href {\doibase
  10.1103/PhysRev.130.1616} {\bibfield  {journal} {\bibinfo  {journal} {Phys.
  Rev.}\ }\textbf {\bibinfo {volume} {130}},\ \bibinfo {pages} {1616} (\bibinfo
  {year} {1963})}\BibitemShut {NoStop}%
\bibitem [{Note1()}]{Note1}%
  \BibitemOpen
  \bibinfo {note} {The inverse magnetic susceptibility at zero temperature is:
  \begin {equation*} \label {Eq:kM0} \kappa _{M}^{-1}\left ( T = 0 \right ) =
  \left ( \protect \frac {\partial ^2 \protect \mathcal {E}_0}{\partial
  \protect \mathaccentV {tilde}07E{n}^2} \right )_{\protect \mathaccentV
  {tilde}07E{n} = 0} = \protect \frac {m c_+^2}{n} - \protect \frac {2 m^2}{\pi
  \hbar n^2} \protect \frac {c_-^2 c_+^2}{c_+ + c_-} \end {equation*} where
  $\protect \mathaccentV {tilde}07E{n} = n_1 - n_2$.}\BibitemShut {Stop}%
\bibitem [{\citenamefont {Stringari}(2009)}]{Stringari2009}%
  \BibitemOpen
  \bibfield  {author} {\bibinfo {author} {\bibfnamefont {S.}~\bibnamefont
  {Stringari}},\ }\bibfield  {title} {\bibinfo {title} {\emph {Density and Spin
  Response Function of a Normal Fermi Gas at Unitarity}},\ }\href {\doibase
  10.1103/PhysRevLett.102.110406} {\bibfield  {journal} {\bibinfo  {journal}
  {Phys. Rev. Lett.}\ }\textbf {\bibinfo {volume} {102}},\ \bibinfo {pages}
  {110406} (\bibinfo {year} {2009})}\BibitemShut {NoStop}%
\bibitem [{\citenamefont {Pitaevskii}\ and\ \citenamefont
  {Stringari}(2016)}]{Pitaevskii2016}%
  \BibitemOpen
  \bibfield  {author} {\bibinfo {author} {\bibfnamefont {L.}~\bibnamefont
  {Pitaevskii}}\ and\ \bibinfo {author} {\bibfnamefont {S.}~\bibnamefont
  {Stringari}},\ }\href@noop {} {\emph {\bibinfo {title} {Bose-Einstein
  Condensation and Superfluidity}}},\ International Series of Monographs on
  Physics\ (\bibinfo  {publisher} {Oxford University Press, Oxford, UK},\
  \bibinfo {year} {2016})\BibitemShut {NoStop}%
\bibitem [{\citenamefont {J\o{}rgensen}\ \emph {et~al.}(2018)\citenamefont
  {J\o{}rgensen}, \citenamefont {Bruun},\ and\ \citenamefont
  {Arlt}}]{Jorgensen2018}%
  \BibitemOpen
  \bibfield  {author} {\bibinfo {author} {\bibfnamefont {N.~B.}\ \bibnamefont
  {J\o{}rgensen}}, \bibinfo {author} {\bibfnamefont {G.~M.}\ \bibnamefont
  {Bruun}}, \ and\ \bibinfo {author} {\bibfnamefont {J.~J.}\ \bibnamefont
  {Arlt}},\ }\bibfield  {title} {\bibinfo {title} {\emph {Dilute Fluid Governed
  by Quantum Fluctuations}},\ }\href {\doibase 10.1103/PhysRevLett.121.173403}
  {\bibfield  {journal} {\bibinfo  {journal} {Phys. Rev. Lett.}\ }\textbf
  {\bibinfo {volume} {121}},\ \bibinfo {pages} {173403} (\bibinfo {year}
  {2018})}\BibitemShut {NoStop}%
\bibitem [{\citenamefont {Landau}\ and\ \citenamefont
  {Lifshitz}(2013{\natexlab{b}})}]{Landau2013}%
  \BibitemOpen
  \bibfield  {author} {\bibinfo {author} {\bibfnamefont {L.~D.}\ \bibnamefont
  {Landau}}\ and\ \bibinfo {author} {\bibfnamefont {E.~M.}\ \bibnamefont
  {Lifshitz}},\ }\href@noop {} {\emph {\bibinfo {title} {Statistical Physics:
  Vol. 5}}}\ (\bibinfo  {publisher} {Elsevier Science, Amsterdam},\ \bibinfo
  {year} {2013})\BibitemShut {NoStop}%
\bibitem [{\citenamefont {Kheruntsyan}\ \emph {et~al.}(2003)\citenamefont
  {Kheruntsyan}, \citenamefont {Gangardt}, \citenamefont {Drummond},\ and\
  \citenamefont {Shlyapnikov}}]{Kheruntsyan2003}%
  \BibitemOpen
  \bibfield  {author} {\bibinfo {author} {\bibfnamefont {K.~V.}\ \bibnamefont
  {Kheruntsyan}}, \bibinfo {author} {\bibfnamefont {D.~M.}\ \bibnamefont
  {Gangardt}}, \bibinfo {author} {\bibfnamefont {P.~D.}\ \bibnamefont
  {Drummond}}, \ and\ \bibinfo {author} {\bibfnamefont {G.~V.}\ \bibnamefont
  {Shlyapnikov}},\ }\bibfield  {title} {\bibinfo {title} {\emph {Pair
  Correlations in a Finite-Temperature 1D Bose Gas}},\ }\href {\doibase
  10.1103/PhysRevLett.91.040403} {\bibfield  {journal} {\bibinfo  {journal}
  {Phys. Rev. Lett.}\ }\textbf {\bibinfo {volume} {91}},\ \bibinfo {pages}
  {040403} (\bibinfo {year} {2003})}\BibitemShut {NoStop}%
\bibitem [{\citenamefont {Astrakharchik}\ and\ \citenamefont
  {Pitaevskii}(2004)}]{Astrakharchik2004}%
  \BibitemOpen
  \bibfield  {author} {\bibinfo {author} {\bibfnamefont {G.~E.}\ \bibnamefont
  {Astrakharchik}}\ and\ \bibinfo {author} {\bibfnamefont {L.~P.}\ \bibnamefont
  {Pitaevskii}},\ }\bibfield  {title} {\bibinfo {title} {\emph {Motion of a
  heavy impurity through a Bose-Einstein condensate}},\ }\href {\doibase
  10.1103/PhysRevA.70.013608} {\bibfield  {journal} {\bibinfo  {journal} {Phys.
  Rev. A}\ }\textbf {\bibinfo {volume} {70}},\ \bibinfo {pages} {013608}
  (\bibinfo {year} {2004})}\BibitemShut {NoStop}%
\bibitem [{\citenamefont {Ota}\ \emph {et~al.}(2019)\citenamefont {Ota},
  \citenamefont {Giorgini},\ and\ \citenamefont {Stringari}}]{Ota2019}%
  \BibitemOpen
  \bibfield  {author} {\bibinfo {author} {\bibfnamefont {M.}~\bibnamefont
  {Ota}}, \bibinfo {author} {\bibfnamefont {S.}~\bibnamefont {Giorgini}}, \
  and\ \bibinfo {author} {\bibfnamefont {S.}~\bibnamefont {Stringari}},\
  }\bibfield  {title} {\bibinfo {title} {\emph {Magnetic Phase Transition in a
  Mixture of Two Interacting Superfluid Bose Gases at Finite Temperature}},\
  }\href {\doibase 10.1103/PhysRevLett.123.075301} {\bibfield  {journal}
  {\bibinfo  {journal} {Phys. Rev. Lett.}\ }\textbf {\bibinfo {volume} {123}},\
  \bibinfo {pages} {075301} (\bibinfo {year} {2019})}\BibitemShut {NoStop}%
\bibitem [{\citenamefont {Hryhorchak}\ and\ \citenamefont
  {Pastukhov}(2021)}]{Hryhorchak2021}%
  \BibitemOpen
  \bibfield  {author} {\bibinfo {author} {\bibfnamefont {O.}~\bibnamefont
  {Hryhorchak}}\ and\ \bibinfo {author} {\bibfnamefont {V.}~\bibnamefont
  {Pastukhov}},\ }\bibfield  {title} {\bibinfo {title} {\emph {Large-N
  Expansion for Condensation and Stability of Bose-Bose Mixtures at Finite
  Temperatures}},\ }\href {https://doi.org/10.1007/s10909-020-02542-y}
  {\bibfield  {journal} {\bibinfo  {journal} {J Low Temp Phys}\ }\textbf
  {\bibinfo {volume} {202}},\ \bibinfo {pages} {219} (\bibinfo {year}
  {2021})}\BibitemShut {NoStop}%
\bibitem [{\citenamefont {Suthar}\ and\ \citenamefont
  {Angom}(2017)}]{Suthar2017}%
  \BibitemOpen
  \bibfield  {author} {\bibinfo {author} {\bibfnamefont {K.}~\bibnamefont
  {Suthar}}\ and\ \bibinfo {author} {\bibfnamefont {D.}~\bibnamefont {Angom}},\
  }\bibfield  {title} {\bibinfo {title} {\emph {Characteristic temperature for
  the immiscible-miscible transition of binary condensates in optical
  lattices}},\ }\href {\doibase 10.1103/PhysRevA.95.043602} {\bibfield
  {journal} {\bibinfo  {journal} {Phys. Rev. A}\ }\textbf {\bibinfo {volume}
  {95}},\ \bibinfo {pages} {043602} (\bibinfo {year} {2017})}\BibitemShut
  {NoStop}%
\bibitem [{\citenamefont {Mora}\ and\ \citenamefont {Castin}(2003)}]{Mora2003}%
  \BibitemOpen
  \bibfield  {author} {\bibinfo {author} {\bibfnamefont {C.}~\bibnamefont
  {Mora}}\ and\ \bibinfo {author} {\bibfnamefont {Y.}~\bibnamefont {Castin}},\
  }\bibfield  {title} {\bibinfo {title} {\emph {Extension of Bogoliubov theory
  to quasicondensates}},\ }\href {\doibase 10.1103/PhysRevA.67.053615}
  {\bibfield  {journal} {\bibinfo  {journal} {Phys. Rev. A}\ }\textbf {\bibinfo
  {volume} {67}},\ \bibinfo {pages} {053615} (\bibinfo {year}
  {2003})}\BibitemShut {NoStop}%
\bibitem [{\citenamefont {Wenzel}\ \emph {et~al.}(2018)\citenamefont {Wenzel},
  \citenamefont {Pfau},\ and\ \citenamefont {Ferrier-Barbut}}]{Wenzel2018}%
  \BibitemOpen
  \bibfield  {author} {\bibinfo {author} {\bibfnamefont {M.}~\bibnamefont
  {Wenzel}}, \bibinfo {author} {\bibfnamefont {T.}~\bibnamefont {Pfau}}, \ and\
  \bibinfo {author} {\bibfnamefont {I.}~\bibnamefont {Ferrier-Barbut}},\
  }\bibfield  {title} {\bibinfo {title} {\emph {A fermionic impurity in a
  dipolar quantum droplet}},\ }\href {\doibase 10.1088/1402-4896/aadd72}
  {\bibfield  {journal} {\bibinfo  {journal} {Phys. Scripta}\ }\textbf
  {\bibinfo {volume} {93}},\ \bibinfo {pages} {104004} (\bibinfo {year}
  {2018})}\BibitemShut {NoStop}%
\bibitem [{\citenamefont {Toennies}\ and\ \citenamefont
  {Vilesov}(2004)}]{Toennies2004}%
  \BibitemOpen
  \bibfield  {author} {\bibinfo {author} {\bibfnamefont {J.~P.}\ \bibnamefont
  {Toennies}}\ and\ \bibinfo {author} {\bibfnamefont {A.~F.}\ \bibnamefont
  {Vilesov}},\ }\bibfield  {title} {\bibinfo {title} {\emph {Superfluid Helium
  Droplets: A Uniquely Cold Nanomatrix for Molecules and Molecular
  Complexes}},\ }\href {\doibase 10.1002/anie.200300611} {\bibfield  {journal}
  {\bibinfo  {journal} {Angewandte Chemie International Edition}\ }\textbf
  {\bibinfo {volume} {43}},\ \bibinfo {pages} {2622} (\bibinfo {year}
  {2004})}\BibitemShut {NoStop}%
\bibitem [{\citenamefont {Hartke}\ \emph {et~al.}(2020)\citenamefont {Hartke},
  \citenamefont {Oreg}, \citenamefont {Jia},\ and\ \citenamefont
  {Zwierlein}}]{Hartke2020}%
  \BibitemOpen
  \bibfield  {author} {\bibinfo {author} {\bibfnamefont {T.}~\bibnamefont
  {Hartke}}, \bibinfo {author} {\bibfnamefont {B.}~\bibnamefont {Oreg}},
  \bibinfo {author} {\bibfnamefont {N.}~\bibnamefont {Jia}}, \ and\ \bibinfo
  {author} {\bibfnamefont {M.}~\bibnamefont {Zwierlein}},\ }\bibfield  {title}
  {\bibinfo {title} {\emph {Doublon-Hole Correlations and Fluctuation
  Thermometry in a Fermi-Hubbard Gas}},\ }\href {\doibase
  10.1103/PhysRevLett.125.113601} {\bibfield  {journal} {\bibinfo  {journal}
  {Phys. Rev. Lett.}\ }\textbf {\bibinfo {volume} {125}},\ \bibinfo {pages}
  {113601} (\bibinfo {year} {2020})}\BibitemShut {NoStop}%
\bibitem [{\citenamefont {De~Daniloff}\ \emph {et~al.}(2021)\citenamefont
  {De~Daniloff}, \citenamefont {Tharrault}, \citenamefont {Enesa},
  \citenamefont {Salomon}, \citenamefont {Chevy}, \citenamefont {Reimann},\
  and\ \citenamefont {Struck}}]{DeDaniloff2021}%
  \BibitemOpen
  \bibfield  {author} {\bibinfo {author} {\bibfnamefont {C.}~\bibnamefont
  {De~Daniloff}}, \bibinfo {author} {\bibfnamefont {M.}~\bibnamefont
  {Tharrault}}, \bibinfo {author} {\bibfnamefont {C.}~\bibnamefont {Enesa}},
  \bibinfo {author} {\bibfnamefont {C.}~\bibnamefont {Salomon}}, \bibinfo
  {author} {\bibfnamefont {F.}~\bibnamefont {Chevy}}, \bibinfo {author}
  {\bibfnamefont {T.}~\bibnamefont {Reimann}}, \ and\ \bibinfo {author}
  {\bibfnamefont {J.}~\bibnamefont {Struck}},\ }\bibfield  {title} {\bibinfo
  {title} {\emph {In Situ Thermometry of Fermionic Cold-Atom Quantum Wires}},\
  }\href {https://arxiv.org/abs/2102.01589} {\bibfield  {journal} {\bibinfo
  {journal} {arXiv:2102.01589}\ } (\bibinfo {year} {2021})}\BibitemShut
  {NoStop}%
\bibitem [{\citenamefont {Desbuquois}\ \emph {et~al.}(2014)\citenamefont
  {Desbuquois}, \citenamefont {Yefsah}, \citenamefont {Chomaz}, \citenamefont
  {Weitenberg}, \citenamefont {Corman}, \citenamefont {Nascimb\`ene},\ and\
  \citenamefont {Dalibard}}]{Desbuquois2014}%
  \BibitemOpen
  \bibfield  {author} {\bibinfo {author} {\bibfnamefont {R.}~\bibnamefont
  {Desbuquois}}, \bibinfo {author} {\bibfnamefont {T.}~\bibnamefont {Yefsah}},
  \bibinfo {author} {\bibfnamefont {L.}~\bibnamefont {Chomaz}}, \bibinfo
  {author} {\bibfnamefont {C.}~\bibnamefont {Weitenberg}}, \bibinfo {author}
  {\bibfnamefont {L.}~\bibnamefont {Corman}}, \bibinfo {author} {\bibfnamefont
  {S.}~\bibnamefont {Nascimb\`ene}}, \ and\ \bibinfo {author} {\bibfnamefont
  {J.}~\bibnamefont {Dalibard}},\ }\bibfield  {title} {\bibinfo {title} {\emph
  {Determination of Scale-Invariant Equations of State without Fitting
  Parameters: Application to the Two-Dimensional Bose Gas Across the
  Berezinskii-Kosterlitz-Thouless Transition}},\ }\href {\doibase
  10.1103/PhysRevLett.113.020404} {\bibfield  {journal} {\bibinfo  {journal}
  {Phys. Rev. Lett.}\ }\textbf {\bibinfo {volume} {113}},\ \bibinfo {pages}
  {020404} (\bibinfo {year} {2014})}\BibitemShut {NoStop}%
\bibitem [{\citenamefont {Zhou}\ \emph {et~al.}(2019)\citenamefont {Zhou},
  \citenamefont {Yu}, \citenamefont {Zou},\ and\ \citenamefont
  {Zhong}}]{Zhou2019}%
  \BibitemOpen
  \bibfield  {author} {\bibinfo {author} {\bibfnamefont {Z.}~\bibnamefont
  {Zhou}}, \bibinfo {author} {\bibfnamefont {X.}~\bibnamefont {Yu}}, \bibinfo
  {author} {\bibfnamefont {Y.}~\bibnamefont {Zou}}, \ and\ \bibinfo {author}
  {\bibfnamefont {H.}~\bibnamefont {Zhong}},\ }\bibfield  {title} {\bibinfo
  {title} {\emph {Dynamics of quantum droplets in a one-dimensional optical
  lattice}},\ }\href {\doibase https://doi.org/10.1016/j.cnsns.2019.104881}
  {\bibfield  {journal} {\bibinfo  {journal} {Comm. Nonlin. Sci. Numer.
  Simul.}\ }\textbf {\bibinfo {volume} {78}},\ \bibinfo {pages} {104881}
  (\bibinfo {year} {2019})}\BibitemShut {NoStop}%
\bibitem [{\citenamefont {Morera}\ \emph {et~al.}(2020)\citenamefont {Morera},
  \citenamefont {Astrakharchik}, \citenamefont {Polls},\ and\ \citenamefont
  {Jul\'ia-D\'iaz}}]{Morera2020}%
  \BibitemOpen
  \bibfield  {author} {\bibinfo {author} {\bibfnamefont {I.}~\bibnamefont
  {Morera}}, \bibinfo {author} {\bibfnamefont {G.~E.}\ \bibnamefont
  {Astrakharchik}}, \bibinfo {author} {\bibfnamefont {A.}~\bibnamefont
  {Polls}}, \ and\ \bibinfo {author} {\bibfnamefont {B.}~\bibnamefont
  {Jul\'ia-D\'iaz}},\ }\bibfield  {title} {\bibinfo {title} {\emph {Quantum
  droplets of bosonic mixtures in a one-dimensional optical lattice}},\ }\href
  {\doibase 10.1103/PhysRevResearch.2.022008} {\bibfield  {journal} {\bibinfo
  {journal} {Phys. Rev. Res.}\ }\textbf {\bibinfo {volume} {2}},\ \bibinfo
  {pages} {022008(R)} (\bibinfo {year} {2020})}\BibitemShut {NoStop}%
\bibitem [{\citenamefont {Morera}\ \emph {et~al.}(2021)\citenamefont {Morera},
  \citenamefont {Astrakharchik}, \citenamefont {Polls},\ and\ \citenamefont
  {Juli\'a-D\'{\i}az}}]{Morera20202}%
  \BibitemOpen
  \bibfield  {author} {\bibinfo {author} {\bibfnamefont {I.}~\bibnamefont
  {Morera}}, \bibinfo {author} {\bibfnamefont {G.~E.}\ \bibnamefont
  {Astrakharchik}}, \bibinfo {author} {\bibfnamefont {A.}~\bibnamefont
  {Polls}}, \ and\ \bibinfo {author} {\bibfnamefont {B.}~\bibnamefont
  {Juli\'a-D\'{\i}az}},\ }\bibfield  {title} {\bibinfo {title} {\emph
  {Universal Dimerized Quantum Droplets in a One-Dimensional Lattice}},\ }\href
  {\doibase 10.1103/PhysRevLett.126.023001} {\bibfield  {journal} {\bibinfo
  {journal} {Phys. Rev. Lett.}\ }\textbf {\bibinfo {volume} {126}},\ \bibinfo
  {pages} {023001} (\bibinfo {year} {2021})}\BibitemShut {NoStop}%
\bibitem [{\citenamefont {Pricoupenko}\ and\ \citenamefont
  {Petrov}(2018)}]{Pricoupenko2018}%
  \BibitemOpen
  \bibfield  {author} {\bibinfo {author} {\bibfnamefont {A.}~\bibnamefont
  {Pricoupenko}}\ and\ \bibinfo {author} {\bibfnamefont {D.~S.}\ \bibnamefont
  {Petrov}},\ }\bibfield  {title} {\bibinfo {title} {\emph {Dimer-dimer zero
  crossing and dilute dimerized liquid in a one-dimensional mixture}},\ }\href
  {\doibase 10.1103/PhysRevA.97.063616} {\bibfield  {journal} {\bibinfo
  {journal} {Phys. Rev. A}\ }\textbf {\bibinfo {volume} {97}},\ \bibinfo
  {pages} {063616} (\bibinfo {year} {2018})}\BibitemShut {NoStop}%
\bibitem [{\citenamefont {Guijarro}\ \emph {et~al.}(2018)\citenamefont
  {Guijarro}, \citenamefont {Pricoupenko}, \citenamefont {Astrakharchik},
  \citenamefont {Boronat},\ and\ \citenamefont {Petrov}}]{Guijarro2018}%
  \BibitemOpen
  \bibfield  {author} {\bibinfo {author} {\bibfnamefont {G.}~\bibnamefont
  {Guijarro}}, \bibinfo {author} {\bibfnamefont {A.}~\bibnamefont
  {Pricoupenko}}, \bibinfo {author} {\bibfnamefont {G.~E.}\ \bibnamefont
  {Astrakharchik}}, \bibinfo {author} {\bibfnamefont {J.}~\bibnamefont
  {Boronat}}, \ and\ \bibinfo {author} {\bibfnamefont {D.~S.}\ \bibnamefont
  {Petrov}},\ }\bibfield  {title} {\bibinfo {title} {\emph {One-dimensional
  three-boson problem with two- and three-body interactions}},\ }\href
  {\doibase 10.1103/PhysRevA.97.061605} {\bibfield  {journal} {\bibinfo
  {journal} {Phys. Rev. A}\ }\textbf {\bibinfo {volume} {97}},\ \bibinfo
  {pages} {061605(R)} (\bibinfo {year} {2018})}\BibitemShut {NoStop}%
\bibitem [{\citenamefont {Cappellaro}\ \emph {et~al.}(2017)\citenamefont
  {Cappellaro}, \citenamefont {Macr\'{i}}, \citenamefont {Bertacco},\ and\
  \citenamefont {Salasnich}}]{Cappellaro2017}%
  \BibitemOpen
  \bibfield  {author} {\bibinfo {author} {\bibfnamefont {A.}~\bibnamefont
  {Cappellaro}}, \bibinfo {author} {\bibfnamefont {T.}~\bibnamefont
  {Macr\'{i}}}, \bibinfo {author} {\bibfnamefont {G.~F.}\ \bibnamefont
  {Bertacco}}, \ and\ \bibinfo {author} {\bibfnamefont {L.}~\bibnamefont
  {Salasnich}},\ }\bibfield  {title} {\bibinfo {title} {\emph {Equation of
  state and self-bound droplet in Rabi-coupled Bose mixtures}},\ }\href
  {\doibase https://doi.org/10.1038/s41598-017-13647-y} {\bibfield  {journal}
  {\bibinfo  {journal} {Sci. Rep.}\ }\textbf {\bibinfo {volume} {7}},\ \bibinfo
  {pages} {13358} (\bibinfo {year} {2017})}\BibitemShut {NoStop}%
\bibitem [{\citenamefont {Bardeen}\ \emph {et~al.}(1967)\citenamefont
  {Bardeen}, \citenamefont {Baym},\ and\ \citenamefont {Pines}}]{Bardeen1967}%
  \BibitemOpen
  \bibfield  {author} {\bibinfo {author} {\bibfnamefont {J.}~\bibnamefont
  {Bardeen}}, \bibinfo {author} {\bibfnamefont {G.}~\bibnamefont {Baym}}, \
  and\ \bibinfo {author} {\bibfnamefont {D.}~\bibnamefont {Pines}},\ }\bibfield
   {title} {\bibinfo {title} {\emph {Effective Interaction of
  ${\mathrm{He}}^{3}$ Atoms in Dilute Solutions of ${\mathrm{He}}^{3}$ in
  ${\mathrm{He}}^{4}$ at Low Temperatures}},\ }\href {\doibase
  10.1103/PhysRev.156.207} {\bibfield  {journal} {\bibinfo  {journal} {Phys.
  Rev.}\ }\textbf {\bibinfo {volume} {156}},\ \bibinfo {pages} {207} (\bibinfo
  {year} {1967})}\BibitemShut {NoStop}%
\bibitem [{\citenamefont {Stienkemeier}\ and\ \citenamefont
  {Lehmann}(2006)}]{Stienkemeier2006}%
  \BibitemOpen
  \bibfield  {author} {\bibinfo {author} {\bibfnamefont {F.}~\bibnamefont
  {Stienkemeier}}\ and\ \bibinfo {author} {\bibfnamefont {K.~K.}\ \bibnamefont
  {Lehmann}},\ }\bibfield  {title} {\bibinfo {title} {\emph {Spectroscopy and
  dynamics in helium nanodroplets}},\ }\href {\doibase
  10.1088/0953-4075/39/8/r01} {\bibfield  {journal} {\bibinfo  {journal} {J.
  Phys. B: At. Mol. Opt. Phys.}\ }\textbf {\bibinfo {volume} {39}},\ \bibinfo
  {pages} {R127} (\bibinfo {year} {2006})}\BibitemShut {NoStop}%
\bibitem [{\citenamefont {Bisset}\ \emph {et~al.}(2021)\citenamefont {Bisset},
  \citenamefont {Pe\~{n}a Ardila},\ and\ \citenamefont {Santos}}]{Bisset2021}%
  \BibitemOpen
  \bibfield  {author} {\bibinfo {author} {\bibfnamefont {R.~N.}\ \bibnamefont
  {Bisset}}, \bibinfo {author} {\bibfnamefont {L.~A.}\ \bibnamefont {Pe\~{n}a
  Ardila}}, \ and\ \bibinfo {author} {\bibfnamefont {L.}~\bibnamefont
  {Santos}},\ }\bibfield  {title} {\bibinfo {title} {\emph {Quantum Droplets of
  Dipolar Mixtures}},\ }\href {\doibase 10.1103/PhysRevLett.126.025301}
  {\bibfield  {journal} {\bibinfo  {journal} {Phys. Rev. Lett.}\ }\textbf
  {\bibinfo {volume} {126}},\ \bibinfo {pages} {025301} (\bibinfo {year}
  {2021})}\BibitemShut {NoStop}%
\bibitem [{\citenamefont {Recati}\ \emph {et~al.}(2005)\citenamefont {Recati},
  \citenamefont {Fuchs}, \citenamefont {Peca},\ and\ \citenamefont
  {Zwerger}}]{Recati2005}%
  \BibitemOpen
  \bibfield  {author} {\bibinfo {author} {\bibfnamefont {A.}~\bibnamefont
  {Recati}}, \bibinfo {author} {\bibfnamefont {J.~N.}\ \bibnamefont {Fuchs}},
  \bibinfo {author} {\bibfnamefont {C.~S.}\ \bibnamefont {Peca}}, \ and\
  \bibinfo {author} {\bibfnamefont {W.}~\bibnamefont {Zwerger}},\ }\bibfield
  {title} {\bibinfo {title} {\emph {Casimir forces between defects in
  one-dimensional quantum liquids}},\ }\href {\doibase
  10.1103/PhysRevA.72.023616} {\bibfield  {journal} {\bibinfo  {journal} {Phys.
  Rev. A}\ }\textbf {\bibinfo {volume} {72}},\ \bibinfo {pages} {023616}
  (\bibinfo {year} {2005})}\BibitemShut {NoStop}%
\bibitem [{\citenamefont {Schecter}\ and\ \citenamefont
  {Kamenev}(2014)}]{Schecter2014}%
  \BibitemOpen
  \bibfield  {author} {\bibinfo {author} {\bibfnamefont {M.}~\bibnamefont
  {Schecter}}\ and\ \bibinfo {author} {\bibfnamefont {A.}~\bibnamefont
  {Kamenev}},\ }\bibfield  {title} {\bibinfo {title} {\emph {Phonon-Mediated
  Casimir Interaction between Mobile Impurities in One-Dimensional Quantum
  Liquids}},\ }\href {\doibase 10.1103/PhysRevLett.112.155301} {\bibfield
  {journal} {\bibinfo  {journal} {Phys. Rev. Lett.}\ }\textbf {\bibinfo
  {volume} {112}},\ \bibinfo {pages} {155301} (\bibinfo {year}
  {2014})}\BibitemShut {NoStop}%
\bibitem [{\citenamefont {Ferioli}\ \emph {et~al.}(2020)\citenamefont
  {Ferioli}, \citenamefont {Semeghini}, \citenamefont {Terradas-Brians\'o},
  \citenamefont {Masi}, \citenamefont {Fattori},\ and\ \citenamefont
  {Modugno}}]{Ferioli2020}%
  \BibitemOpen
  \bibfield  {author} {\bibinfo {author} {\bibfnamefont {G.}~\bibnamefont
  {Ferioli}}, \bibinfo {author} {\bibfnamefont {G.}~\bibnamefont {Semeghini}},
  \bibinfo {author} {\bibfnamefont {S.}~\bibnamefont {Terradas-Brians\'o}},
  \bibinfo {author} {\bibfnamefont {L.}~\bibnamefont {Masi}}, \bibinfo {author}
  {\bibfnamefont {M.}~\bibnamefont {Fattori}}, \ and\ \bibinfo {author}
  {\bibfnamefont {M.}~\bibnamefont {Modugno}},\ }\bibfield  {title} {\bibinfo
  {title} {\emph {Dynamical formation of quantum droplets in a
  $^{39}\mathrm{K}$ mixture}},\ }\href {\doibase
  10.1103/PhysRevResearch.2.013269} {\bibfield  {journal} {\bibinfo  {journal}
  {Phys. Rev. Res.}\ }\textbf {\bibinfo {volume} {2}},\ \bibinfo {pages}
  {013269} (\bibinfo {year} {2020})}\BibitemShut {NoStop}%
\bibitem [{\citenamefont {Ferioli}\ \emph {et~al.}(2019)\citenamefont
  {Ferioli}, \citenamefont {Semeghini}, \citenamefont {Masi}, \citenamefont
  {Giusti}, \citenamefont {Modugno}, \citenamefont {Inguscio}, \citenamefont
  {Gallem\'{\i}}, \citenamefont {Recati},\ and\ \citenamefont
  {Fattori}}]{Ferioli2019}%
  \BibitemOpen
  \bibfield  {author} {\bibinfo {author} {\bibfnamefont {G.}~\bibnamefont
  {Ferioli}}, \bibinfo {author} {\bibfnamefont {G.}~\bibnamefont {Semeghini}},
  \bibinfo {author} {\bibfnamefont {L.}~\bibnamefont {Masi}}, \bibinfo {author}
  {\bibfnamefont {G.}~\bibnamefont {Giusti}}, \bibinfo {author} {\bibfnamefont
  {G.}~\bibnamefont {Modugno}}, \bibinfo {author} {\bibfnamefont
  {M.}~\bibnamefont {Inguscio}}, \bibinfo {author} {\bibfnamefont
  {A.}~\bibnamefont {Gallem\'{\i}}}, \bibinfo {author} {\bibfnamefont
  {A.}~\bibnamefont {Recati}}, \ and\ \bibinfo {author} {\bibfnamefont
  {M.}~\bibnamefont {Fattori}},\ }\bibfield  {title} {\bibinfo {title} {\emph
  {Collisions of Self-Bound Quantum Droplets}},\ }\href {\doibase
  10.1103/PhysRevLett.122.090401} {\bibfield  {journal} {\bibinfo  {journal}
  {Phys. Rev. Lett.}\ }\textbf {\bibinfo {volume} {122}},\ \bibinfo {pages}
  {090401} (\bibinfo {year} {2019})}\BibitemShut {NoStop}%
\bibitem [{\citenamefont {D'Errico}\ \emph {et~al.}(2019)\citenamefont
  {D'Errico}, \citenamefont {Burchianti}, \citenamefont {Prevedelli},
  \citenamefont {Salasnich}, \citenamefont {Ancilotto}, \citenamefont
  {Modugno}, \citenamefont {Minardi},\ and\ \citenamefont
  {Fort}}]{DErrico2019}%
  \BibitemOpen
  \bibfield  {author} {\bibinfo {author} {\bibfnamefont {C.}~\bibnamefont
  {D'Errico}}, \bibinfo {author} {\bibfnamefont {A.}~\bibnamefont
  {Burchianti}}, \bibinfo {author} {\bibfnamefont {M.}~\bibnamefont
  {Prevedelli}}, \bibinfo {author} {\bibfnamefont {L.}~\bibnamefont
  {Salasnich}}, \bibinfo {author} {\bibfnamefont {F.}~\bibnamefont
  {Ancilotto}}, \bibinfo {author} {\bibfnamefont {M.}~\bibnamefont {Modugno}},
  \bibinfo {author} {\bibfnamefont {F.}~\bibnamefont {Minardi}}, \ and\
  \bibinfo {author} {\bibfnamefont {C.}~\bibnamefont {Fort}},\ }\bibfield
  {title} {\bibinfo {title} {\emph {Observation of quantum droplets in a
  heteronuclear bosonic mixture}},\ }\href {\doibase
  10.1103/PhysRevResearch.1.033155} {\bibfield  {journal} {\bibinfo  {journal}
  {Phys. Rev. Res.}\ }\textbf {\bibinfo {volume} {1}},\ \bibinfo {pages}
  {033155} (\bibinfo {year} {2019})}\BibitemShut {NoStop}%
\bibitem [{\citenamefont {Burchianti}\ \emph {et~al.}(2020)\citenamefont
  {Burchianti}, \citenamefont {D'Errico}, \citenamefont {Prevedelli},
  \citenamefont {Salasnich}, \citenamefont {Ancilotto}, \citenamefont
  {Modugno}, \citenamefont {Minardi},\ and\ \citenamefont
  {Fort}}]{Burchianti2020}%
  \BibitemOpen
  \bibfield  {author} {\bibinfo {author} {\bibfnamefont {A.}~\bibnamefont
  {Burchianti}}, \bibinfo {author} {\bibfnamefont {C.}~\bibnamefont
  {D'Errico}}, \bibinfo {author} {\bibfnamefont {M.}~\bibnamefont
  {Prevedelli}}, \bibinfo {author} {\bibfnamefont {L.}~\bibnamefont
  {Salasnich}}, \bibinfo {author} {\bibfnamefont {F.}~\bibnamefont
  {Ancilotto}}, \bibinfo {author} {\bibfnamefont {M.}~\bibnamefont {Modugno}},
  \bibinfo {author} {\bibfnamefont {F.}~\bibnamefont {Minardi}}, \ and\
  \bibinfo {author} {\bibfnamefont {C.}~\bibnamefont {Fort}},\ }\bibfield
  {title} {\bibinfo {title} {\emph {A Dual-Species Bose-Einstein Condensate
  with Attractive Interspecies Interactions}},\ }\href {\doibase
  https://doi.org/10.3390/condmat5010021} {\bibfield  {journal} {\bibinfo
  {journal} {Condens. Matter}\ }\textbf {\bibinfo {volume} {5}},\ \bibinfo
  {pages} {21} (\bibinfo {year} {2020})}\BibitemShut {NoStop}%
\bibitem [{\citenamefont {Mithun}\ \emph {et~al.}(2020)\citenamefont {Mithun},
  \citenamefont {Maluckov}, \citenamefont {Kasamatsu}, \citenamefont
  {Malomed},\ and\ \citenamefont {Khare}}]{Mithun2020}%
  \BibitemOpen
  \bibfield  {author} {\bibinfo {author} {\bibfnamefont {T.}~\bibnamefont
  {Mithun}}, \bibinfo {author} {\bibfnamefont {A.}~\bibnamefont {Maluckov}},
  \bibinfo {author} {\bibfnamefont {K.}~\bibnamefont {Kasamatsu}}, \bibinfo
  {author} {\bibfnamefont {B.}~\bibnamefont {Malomed}}, \ and\ \bibinfo
  {author} {\bibfnamefont {A.}~\bibnamefont {Khare}},\ }\bibfield  {title}
  {\bibinfo {title} {\emph {Modulational Instability, Inter-Component
  Asymmetry, and Formation of Quantum Droplets in One-Dimensional Binary Bose
  Gases}},\ }\href {\doibase https://doi.org/10.3390/sym12010174} {\bibfield
  {journal} {\bibinfo  {journal} {Symmetry}\ }\textbf {\bibinfo {volume}
  {12}},\ \bibinfo {pages} {174} (\bibinfo {year} {2020})}\BibitemShut
  {NoStop}%
\bibitem [{\citenamefont {Naidon}\ and\ \citenamefont
  {Petrov}(2021)}]{Naidon2021}%
  \BibitemOpen
  \bibfield  {author} {\bibinfo {author} {\bibfnamefont {P.}~\bibnamefont
  {Naidon}}\ and\ \bibinfo {author} {\bibfnamefont {D.~S.}\ \bibnamefont
  {Petrov}},\ }\bibfield  {title} {\bibinfo {title} {\emph {Mixed bubbles in
  Bose-Bose mixtures}},\ }\href
  {https://journals.aps.org/prl/abstract/10.1103/PhysRevLett.126.115301}
  {\bibfield  {journal} {\bibinfo  {journal} {Phys. Rev. Lett.}\ }\textbf
  {\bibinfo {volume} {126}},\ \bibinfo {pages} {115301} (\bibinfo {year}
  {2021})}\BibitemShut {NoStop}%
\end{thebibliography}%

\end{document}